\RequirePackage{fix-cm}
\documentclass[twocolumn]{svjour3}          % twocolumn
\smartqed  % flush right qed marks, e.g. at end of proof

\usepackage{graphicx}
\usepackage{graphicx}
\usepackage{empheq}
\usepackage{subfigure}

\usepackage{tikz-cd}
\usepackage{tikz}

\usepackage[numbers]{natbib}
\usepackage[integrals]{wasysym}
\usepackage{changes}

\usetikzlibrary{decorations.markings}
\usetikzlibrary{shapes.geometric, arrows}
\usepackage{tikzscale}
\usepackage{filecontents}
\usepackage{wrapfig}
\usepackage{tcolorbox}
\usepackage{color}
\definecolor{blue1}{rgb}{0.0, 0.0, 1.0}
\definecolor{gray}{rgb}{0.9,0.9,0.9}
\definecolor{gray1}{rgb}{0.7,0.7,0.7}
\definecolor{gray2}{rgb}{0.8,0.8,0.8}
\definecolor{magenta}{rgb}{1.0, 0.0, 1.0}
\usepackage[title]{appendix}
\newcommand\backmatter{\appendix
\def\chaptermark##1{\markboth{%
\ifnum  \c@secnumdepth > \m@ne  \@chapapp\ \thechapter:  \fi  ##1}{%
\ifnum  \c@secnumdepth > \m@ne  \@chapapp\ \thechapter:  \fi  ##1}}%
\def\sectionmark##1{\relax}}

\newcommand{\ISI}[1]{\mathrm{ISI}}

\usepackage{hyperref}
\hypersetup{ colorlinks=true, urlcolor  = blue, linkcolor = blue,
citecolor = blue1,}

\usepackage{float}
\usepackage{amsmath,mathtools,amssymb,amsfonts}
\usepackage[scaled=.95]{helvet}

\definecolor{lime}{HTML}{A6CE39}
\DeclareRobustCommand{\orcidicon}{%
    \begin{tikzpicture}
    \draw[lime, fill=lime] (0,0) 
    circle [radius=0.16] 
    node[white] {{\fontfamily{qag}\selectfont \tiny ID}};
    \draw[white, fill=white] (-0.0625,0.095) 
    circle [radius=0.007];
    \end{tikzpicture}
    \hspace{-2mm}
}

\foreach \x in {A, ..., Z}{%
    \expandafter\xdef\csname orcid\x\endcsname{\noexpand\href{https://orcid.org/\csname orcidauthor\x\endcsname}{\noexpand\orcidicon}}
}
\newcommand{\orcid}[1]{\href{https://orcid.org/#1}{\textcolor[HTML]{A6CE39}{\orcidicon}}}

\newcommand{\uproman}[1]{\uppercase\expandafter{\romannumeral#1}}

 \journalname{Journal}

\begin{document}
\title{Coherence resonance and stochastic synchronization in a small-world neural network: An interplay in the presence of spike-timing-dependent plasticity}

\titlerunning{Coherence resonance and stochastic synchronization in a small-world neural network}        % if too long for running head

\author{Marius E. Yamakou\orcid{0000-0002-2809-1739},  Estelle M. Inack\orcid{0000-0002-4672-5512}}


\institute{M. E. Yamakou \at Department of Data Science, Friedrich-Alexander-Universit\"at Erlangen-Nürnberg, Cauerstr. 11, 91058 Erlangen, Germany.
\at Max-Planck-Institut f\"ur Mathematik in den Naturwissenschaften, Inselstr. 22, 04103 Leipzig, Germany.\\
Estelle M. Inack \at Perimeter Institute for Theoretical Physics, Waterloo, ON N2L 2Y5, Canada.
\at Vector Institute, MaRS Centre, Toronto, Ontario, M5G 1M1, Canada.
\at yiyaniQ, Toronto,  Ontario,  M4V 0A3,  Canada\\
             \email{marius.yamakou@fau.de}\\  
             \email{einack@perimeterinstitute.ca}           %  \\
            %of F. Author  %  if needed
}

\date{Received: date / Accepted: date}

\maketitle
\begin{abstract}
 Coherence resonance (CR), stochastic synchronization (SS), and spike-timing-dependent plasticity (STDP) are ubiquitous dynamical processes in biological neural networks. Whether there exists an optimal network and STDP configuration at which CR and SS are both pronounced is a fundamental question of interest that is still elusive. We expect such a configuration to enable the brain to make synergistic and optimal use of these phenomena to process information efficiently. This paper considers a small-world network of excitable Hodgkin-Huxley neurons driven by channel noise and STDP with an asymmetric Hebbian time window. Numerical results indicate specific network topology and STDP parameter intervals in which CR and SS can be simultaneously enhanced. 
 Our results imply that an optimally tuned inherent background noise, STDP rule, and network topology can play a constructive role in enhancing both the time precision of firing and the synchronization in neural systems. 
\end{abstract}

%%%%%%%%%%%%%%%%%%%%%%%%%%%%%%%%%%%%%%%%%%%%%%%%%%%%%%%%%%%%%%%%%%%%%%%%%%%%%%%%%%%%%%%%%%%%%%%%%%%%%%%%%%%%%%%%%%%%%%%%%
\section{Introduction}\label{Sec. I}
Biological neural networks are stochastic nonlinear dynamical systems that can exhibit a plethora of complex phenomena, including, 
STDP \cite{bi2001synaptic,gerstner1996neuronal}, noise-induced resonance phenomena such as stochastic resonance \cite{longtin1993stochastic,lindner2004effects} and coherence resonance \cite{pikovsky1997coherence,wang2010spatial}, and synchronization phenomena \cite{osipov2007synchronization,arenas2008synchronization}, amongst others.
This paper focuses on the interplay between synaptic plasticity governed by an STDP learning rule, coherence resonance (CR), and stochastic synchronization (SS).
CR occurs when there exists a maximal degree of coherence in the spiking times of the excitable neuron for some intermediate noise intensity \cite{pikovsky1997coherence,neiman1997coherence}.

Several studies have reported the rich dynamical behavior of CR with respect to different aspects of neural networks.
It has been shown that for short synaptic time delay, there always exists an intermediate noise intensity at which spatial CR (i.e., CR in spatially extended systems) is maximal. Furthermore, the authors found that the magnitude of SCR  progressively deteriorates as the fraction of rewired synapses increases, implying that the deterioration of SCR in delayed networks with a small-world topology \cite{wang2010spatial,sun2008spatial}. It has also been shown that CR does not only occur when the noise intensity in the neural system is varied. With a constant noise intensity, CR has been shown to occur when another system parameter, e.g., time delay \cite{wang2008delay}, is varied. It has been found that the adjusting rate parameter of STDP can be used to control the degree of CR in a single layer and multiplex neural networks \cite{yamakou2022optimal}. Thus, in the current paper, we investigate CR and its interplay with synchronization as a function of the STDP and network parameters.

On the other hand, synchronization can be understood as a process where several systems adjust their oscillations or a given property of their motion (e.g., amplitude, phase, velocity, etc.) over time due to their interactions \cite{boccaletti2002synchronization,yamakou2016ratcheting,yamakou2020chaotic}. In neural systems, synchronization can emerge from the interaction between neurons or among neural networks and can have significant consequences on all neurons and network functioning \cite{xu2017synchronization,ma2017phase}. It is well-known that synchronization of neural activity within and across brain regions promotes normal physiological functioning, such as the precise temporal coordination
of processes underlying cognition, memory, and perception \cite{neustadter2016eeg}, but it is also well-known to be responsible for some pathological behaviors such as epilepsy \cite{lehnertz2009synchronization}.

STDP has been shown to play a significant role in neural synchronization. For example, in \cite{nowotny2003enhancement}, it has been experimentally observed that the STDP plasticity curve appears to be designed to adjust synaptic strength to a value suitable for stable entrainment of the postsynaptic neuron, thereby significantly enhancing synchronization. It has also been shown that synaptic plasticity based on burst-timing-dependent plasticity induces a different evolution of chimera states compared to synchronous and asynchronous states \cite{wang2020chimeras}. Interestingly, it has also been shown that as the parameter related to the short-term plasticity increases, different combinations of the synaptic rise and decay time lead to a rich transition of spike propagation. During the transition, it is possible to have the coexistence of simple and composite traveling waves and the coexistence of stable and unstable waves if short-term plasticity is properly configured \cite{zhang2013synaptic}.

Noise-induced coherence resonance phenomena and synchronization of coupled excitable neurons may elucidate how the coherent spontaneously synchronized oscillations, which have been observed in the cortex \cite{wang199740,nicolelis1995sensorimotor} are established. Thus, the study of noise-induced resonance phenomena, synchronization, and their interplay in complex neural networks are of great significance.  An active research topic in theoretical neuroscience is to elucidate the synergetic functional roles of two or more different dynamical phenomena in information processing -- noise-induced resonance phenomena on one hand and synchronization on the other. But we must be cognizant of the fact that the ubiquity of noise (and hence noise-induced resonance phenomena) and synchronization, in addition to the inherently adaptive nature of neurobiological networks, condemn these two categories of complex phenomena to interact and affect each other, with plausible significant consequences on the efficiency of information processing. The literature on noise-induced resonance  and synchronization phenomena is abundant, see, e.g., \cite{lindner2004effects,anishchenko2007nonlinear,neiman1999synchronization,tang2014synchronization} and the references therein. Thus, in sequel of this paper, to be brief and concise, we will only focus on that part of the literature in which we intends to make a contribution, i.e., on the interplay between CR and synchronization in adaptive neural networks driven by STDP.

Many research papers have independently investigated the various types of noise-induced resonance phenomena and synchronization in adaptive neural networks driven by STDP, see, e.g., \cite{yu2015spike,yu2014effects} and the references therein. 
In particular, it is shown in \cite{yu2015spike} that as the adjusting rate parameter of the STDP rule increases, the average coupling strength of the network is weakened. Consequently, the degree of CR and spiking synchronization induced by the noise and the random shortcuts of the Newman-Watts small-world network topology deteriorates. In the case of CR, more connections are needed to optimize the temporal coherence-related random shortcuts with a larger adjusting rate. While for noise-induced synchronization, the synchrony is evoked by a lower value of noise intensity. In \cite{yu2014effects}, using the Fourier coefficient -- a measure of the response of output to input frequency in the presence of an external periodic stimulus -- to quantitatively characterize the degree of stochastic resonance (SR), it is shown that (i) the degree of SR can be largely enhanced by STDP. In particular, the resonance for adaptive coupling can reach a much larger value than that of static (i.e., no STDP) one when the noise intensity is small or intermediate, and (ii) STDP with dominant depression and small temporal window ratio is more efficient in enhancing the degree SR.

A few research papers have investigated the interplay between CR and synchronization in non-adaptive neural networks. 
In \cite{andreev2018coherence}, the relation between CR and complete synchronization is investigated in a non-adaptive network (i.e., a network without STDP) of globally coupled neural oscillators with randomly distributed coupling strengths under the influence of intrinsic noise. It is shown that the highest degree of synchronization coincides with the highest degree of CR only with respect to the network size parameter, and no resonances between synchronization and CR are detected for other parameters, including intrinsic noise, external stimulus, and the number of stimulated neurons. The incoincidence between the highest degree of CR and synchronization in \cite{andreev2018coherence} is explained by the fact that: (i) the neural network considered has two different timescales --- that of spiking and bursting --- and thus, the neurons can be synchronized by bursts, but not necessarily by spikes, whereas the degree of CR is calculated from the inter-spike intervals and not the inter-burst intervals (ii) the neurons with irregular (chaotic) bursting dynamics can be well synchronized, while the degree of CR in the network becomes very low due to irregular oscillations.

In \cite{gong2005optimal}, the interplay between CR and synchronization in a non-adaptive spiking network of electrically coupled neurons is investigated in terms of the fraction of random shortcuts in the network. It is shown that when degree of CR reaches the best level at an optimal value of the random shortcuts, the indicator of the degree of synchronization has already decreased to a relatively low value, i.e., the synchronization is already good. Therefore, the best degree of CR and a relatively good degree of synchronization is attained at an optimal (intermediate) value of the fraction of random shortcuts in the network.

However, the studies on the interplay between CR and synchronization presented in \cite{andreev2018coherence,gong2005optimal} ignore one essential dynamical process that drives biological neural networks --- STDP. The main contribution of this paper is to bridge this gap. The paper focuses on the interplay between CR and SS in Watts-Strogatz small-world networks of Hodgkin-Huxley neurons driven by STDP. In particular, we address the following question: how do the parameters of the network topology and/or those of the STDP learning rule combine to individually and simultaneously improve the degree of CR and SS? It is worth noting that, in this paper, we focused on small-world network topology because it consists of one of the critical properties of brain networks and has been shown to play pivotal roles in the coordination of information flow in neural networks \cite{gosak2022networks,hilgetag2016brain}.

Our numerical results indicate that to simultaneously achieve an enhanced degree of CR and SS, the small-world network requires (i) an intermediate (not too low nor too high) average degree connectivity $\langle k \rangle$ (for the neuron model used we need $1<\langle k \rangle<25$, i.e., $\langle k \rangle_{opt}=9$ to be precise)  and  (ii) a low number of random shortcuts in the network connectivity (i.e., a small-world network built with a low rewiring probability $0<\beta\ll1$).  In addition to the two requirements above, to further enhance the degree of both CR and SS in the small-world network, the STDP rule requires (i) a large potentiation adjusting rate parameter $A_1$ and (ii) a small depression temporal window $\tau_2$.

The remaining of this article adheres to the following sequence: In Sec. \ref{Sec. II}, we present the neural network model and the type of STDP learning rule used in our study. In Sec. \ref{Sec. III}, we describe the computational methods used and the quantities that measure the degree of CR and SS. In Sec. \ref{Sec. IV}, we present and discuss the numerical results. We summarize and conclude our findings in Sec. \ref{Sec. V}.
%%%%%%%%%%%%%%%%%%%%%%%%%%%%%%%%%%%%%%%%%%%%%%%%%%%%%%%%%%%%%%%
\section{Model description}\label{Sec. II}
\subsection{Stochastic Hodgkin-Huxley neural networks}
%%%%%%%%%%%%%%%%%%%%%%%%%%%%%%%%%%%%%%%%%%%%%%%%%%%%%%%%%%%%%%%
We consider a network of Hodgkin-Huxley (HH) neurons \cite{hodgkin1952quantitative}, described by following set of coupled stochastic differential equations:
\begin{eqnarray}\label{eq:1}
\begin{split}
\left\{\begin{array}{lcl}
\displaystyle{C_m\frac{dV_{i}}{dt}} &=&- g_{_{\text{Na}}}^{max}m_{i}^3h_{i}(V_{i}-V_{_{\text{Na}}}) - g_{_{\text{L}}}^{max}(V_{i}-V_{_{\text{L}}})\\ 
\hspace{1.55cm}&-& g_{_{\text{K}}}^{max}n_{i}^4(V_{i}-V_{_{\text{K}}}) + I_{i}^{e} -I_{i}^{syn}(t),\\[1.0mm]
\displaystyle{\frac{dx_{i}}{dt}} &=& \alpha_{{x_{i}}}(V_{i})(1 - x_{i}) - \beta_{{x_{i}}}(V_{i})x_{i} + \displaystyle{\xi_{x_i}(t)},
\end{array}\right.
\end{split}
\end{eqnarray}
where the variable $V_i$, $i=1,...,N,$ represents the membrane potential (measured in $\mathrm{mV}$) of neuron $i$, and $t$ is the time measured in $\mathrm{msec}$.
The capacitance of the membrane of each neuron is represented by $C_m = 1 \mathrm{\mu F/cm^3}$.  $I_{i}^{e}$ (measured in $\mathrm{\mu A/cm^2}$) represents a constant external bias current injected into the $i$th neuron in the network. The conductances $g_{_{\text{Na}}}^{max}=120$ $\mathrm{mS/cm^2}$, $g_{_{\text{K}}}^{max}=36$ $\mathrm{mS/cm^2}$,  and $g_{_{\text{L}}}^{max}=0.3$ $\mathrm{mS/cm^2}$ 
respectively denote the maximal sodium, potassium, and leakage conductance, when all ion channels are open. The potentials $V_{_{\text{Na}}}= 50.0$ $\mathrm{mV}$, $V_{_{\text{K}}} = - 77.0$ $\mathrm{mV}$, and $V_{_{\text{L}}}=-54.4$ $\mathrm{mV}$ are the reversal potentials for sodium,
potassium and leak channels, respectively. 

$x_{i}=\{m_{i},h_{i},n_{i}\}$ represent auxiliary dimensionless $[0,1]$-valued dynamical variables representing  the sodium activation, sodium inactivation, and the potassium activation, respectively.
The dynamics of the gating variables $x_i$, depending on the voltage-dependent opening and closing rate functions $\alpha_{{x_{i}}}(V_i)$ and $\beta_{{x_{i}}}(V_i)$, are given by:
\begin{equation}\label{eq:2}
\begin{split}
\left\{\begin{array}{lcl}
\alpha_{m_i}(V_i) &=& \displaystyle{ \frac{(V_i+40)/10}{1-\exp{[-(V_i+40)/10}]},}\\ [4.0mm]
\beta_{m_i}(V_i) &=&  \displaystyle{4\exp{\big[-(V_i+65)/18\big]},}\\[4.0mm] 
\alpha_{h_i}(V_i) &=&  \displaystyle{0.07\exp{\big[-(V_i+65)/20\big]},}\\[2.0mm] 
\beta_{h_i}(V_i) &=&  \displaystyle{\frac{1}{1+\exp{[-(V_i+35)/10]}},}\\[4.0mm] 
\alpha_{n_i}(V_i) &=&  \displaystyle{\frac{(V_i+55)/100}{1-\exp{[-(V_i+55)/10}]},}\\[4.0mm]
\beta_{n_i}(V_i) &=&  \displaystyle{0.125\exp{\big[-(V_i+65)/80\big]}}.
\end{array}\right.
\end{split}
\end{equation}
$\xi_{x_i}(t)$ in Eq.\eqref{eq:1} represent ion channel noises in the HH neural network. Different computational algorithms have been proposed for channel noise \cite{goldwyn2011and}. In this study, we use the sub-unit noise as the ion channel noises \cite{fox1997stochastic,white2000channel}  where $\xi_{x_i}(t)$ are given by independent zero mean Gaussian white noise sources whose auto-correlation functions are given as below
\begin{equation}\label{eq:3}
\begin{split}
\left\{\begin{array}{lcl}
 \langle\xi_{m_i}(t)\xi_{m_i}(t')\rangle = \displaystyle{ \frac{2\alpha_{m_i}V_i\beta_{m_i}V_i}{N_{Na}\big[\alpha_{m_i}V_i+\beta_{m_i}V_i\big]}\delta(t-t')},\\ [5.0mm]
\langle\xi_{h_i}(t)\xi_{h_i}(t')\rangle = \displaystyle{ \frac{2\alpha_{h_i}V_i\beta_{h_i}V_i}{N_{Na}\big[\alpha_{h_i}V_i+\beta_{h_i}V_i\big]}\delta(t-t')},\\ [5.0mm]
\langle\xi_{n_i}(t)\xi_{n_i}(t')\rangle = \displaystyle{ \frac{2\alpha_{n_i}V_i\beta_{n_i}V_i}{N_{K}\big[\alpha_{n_i}V_i+\beta_{n_i}V_i\big]}\delta(t-t')},
\end{array}\right.
\end{split}
\end{equation}
where $N_{Na}$ and $N_K$ represent the total number of opened sodium and potassium channels within a membrane patch. They are calculated as: $N_{Na} = \rho_{Na}S_i$ and $N_K = \rho_KS_i$, where $\rho_{Na} = 60.0$ $\mathrm{\mu m^{-2}}$ and $\rho_K = 18.0$  $\mathrm{\mu m^{-2}}$ are the sodium and potassium channel densities, respectively. $S_i$ is the membrane patch area (in $\mathrm{\mu m^{2}}$) of the $i$th neuron. For simplicity, we assume that all the neurons are identical with the same membrane patch area, i.e., we chose $S_1=S_2=...=S_N=S$. 

%%%%%%%%%%%%%%%%%%%%%%%%%%%%%%%%%%%%%%%%%%%%%%%%%%%%%%%%%%%%%%%
\subsection{Excitable regime and STDP learning rule}
\begin{figure}
\centering
\includegraphics[width=4.0cm,height=4.0cm]{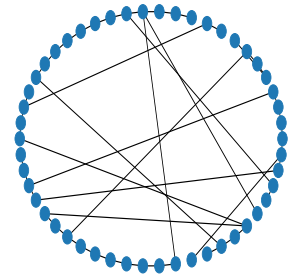}
\caption{Schematic diagram of a Watts-Strogatz small-world network with $N(=50)$ nodes, a rewiring probability of $\beta=0.25$, and an average degree of $\langle k \rangle = 2$.
}
\label{fig:1}
\end{figure}

%%%%%%%%%%%%%%%%%%%%%%%%%%%%%%%%%%%%%%%%%%%%%%%%%%%%%%%%%%%%%%%
The term $I_{i}^{syn}(t)$ in Eq.\eqref{eq:1} models the (inhibitory) chemical synaptic couplings and also governs the STDP learning rule between connected HH neurons. It is given by:
\begin{equation}\label{eq:5}
I_{i}^{syn}(t) = \sum_{\mathclap{j=1(\neq i)}}^{N}l_{ij}g_{ij}(t)s_j(t)\big(V_i(t)- V_{syn}\big),
\end{equation}
where the synaptic connectivity matrix $L (=\{l_{ij}\})$ has $l_{ij}=1$ if the neuron $j$ is pre-synaptic to the neuron $i$; otherwise, $l_{ij}=0$. The synaptic connection is modeled in terms of the time-invariant Watts-Strogatz small-world network \cite{watts1998collective,strogatz2001exploring,watts2000small}. 
The average degree connectivity and rewiring probability (represented by $\langle k \rangle$ and $\beta$, respectively) of a regular network are the parameters of the Watts-Strogatz algorithm \cite{watts1998collective,strogatz2001exploring} which we use to generate our small-world networks. The rewiring probability $\beta\in[0,1]$ represents the fraction of random shortcuts connections added to the regular network (for which $\beta=0$), to form a small-world network when $0<\beta<1$ or a completely random network when $\beta=1$. The rewiring probability $\beta$ and the average degree  $\langle k \rangle$ would be the network parameters used to investigate the interplay between CR and SS. Fig.~\ref{fig:1} shows a schematic representation of a Watts-Strogatz small-world network.

The fraction of open synaptic ion channels at time $t$ of the $j$th neuron is represented by $s_j(t)$ in Eq.\eqref{eq:5} and its time-evolution is governed by \cite{destexhe1994efficient,golomb1993dynamics}:
 \begin{equation}\label{eq:7}
\frac{ds_j}{dt} = \frac{5(1 - s_j)}{1 + \displaystyle{\exp\Big(- \frac{V_j+3}{8}\Big)}}-s_j.
 \end{equation}
 
Luccioli et al.  \cite{luccioli2006dynamical} have investigated the 
three fundamental dynamical regimes of the HH neuron model: (i) an excitable regime in which small values of constant external bias current (i.e., $I^e < 6.27$ $\mathrm{\mu A/cm^2}$) are incapable of initiated or maintaining a self-sustained oscillations of the action potential; (ii) a bistability regime where silence (no spiking) and repetitive firing (limit cycle) co-exist for $6.27 \mathrm{\mu A/cm^2}\le I^e \le 9.78 \mathrm{\mu A/cm^2}$ and depends on the initial conditions; and (iii) repetitive firing regime (stable limit cycle) for $I^e>9.78 \mathrm{\mu A/cm^2}$. Throughout this work,  we fix the constant external bias current at $I^e=6.0$ $\mathrm{\mu A/cm^2}$, in which case the deterministic neural network will be set into a homogeneous excitable regime before channel noise is added into the system to investigate CR and SS in the presence of STDP.

Furthermore, in this work we only consider the kind of synaptic potential that makes a post-synaptic neuron unable to generate a spike due to the occurrence of a spike in the pre-synaptic neuron, i.e, we consider only inhibitory postsynaptic potential (IPSP) \cite{coombs1955specific} and not excitatory postsynaptic potential (EPSP), which is the opposite phenomenon. That is, we want each neuron to spike due solely to the presence of noise and not because it is excited by a spike coming from a the pre-synaptic neuron. Therefore, to have inhibitory chemical synapses (mediated by the GABA$_A$ receptors \cite{brunel2003determines}), we fix, throughout this paper, the reversal synaptic potential at $V_{syn}=-75.0$ $\mathrm{mV}$. It is worth noting that when $V_{syn} < V_i$, the chemical interaction has a depolarizing effect which makes the synapse inhibitory, and when $V_{syn} > V_i$, the chemical interaction has a hyper-polarizing effect, making the synapse excitatory. This means that one can choose $V_{syn}$ such that the inhibitory and excitatory nature of the chemical synapse is determined only by the sign in front of the synaptic current $I_{i}^{syn}(t)$ in Eq.\eqref{eq:1} --- namely, a negative sign.  The negative sign in front of synaptic current $I_{i}^{syn}(t)$ in Eq.\eqref{eq:1} and the relative low value of the external current (i.e., $I^e=6.0$ $\mathrm{\mu A/cm^2}$ which is below the subcritical Hopf bifurcation threshold at $I^e_H=6.27$ $\mathrm{\mu A/cm^2}$) ensure that these chemical synapses are permanently inhibitory. In this setting, any spikes generated in the neurons will be solely induced by their independent channel noise sources and not through the phenomenon of EPSP. Therefore, the spiking activity, its coherence, and its synchronization would be purely noise-induced.
 
The weight of the synaptic connection from the pre-synaptic $j$th neuron to the post-synaptic $i$th neuron is represented by $g_{ij}(t)$ in Eq.\eqref{eq:5}. In this work, we investigate additive STDP, where the coupling update depends on the current value of the synaptic weight $g_{ij}(t)$ and leads to ``hard'' bounds \cite{rubin2001equilibrium,song2000competitive}. With increasing time $t$, the synaptic strength $g_{ij}$ for each synapse is updated with a nearest-spike pair-based STDP rule \cite{morrison2007spike}. To prevent unbounded growth, negative conductances (i.e., negative coupling strength) and elimination of synapses (i.e., $g_{ij}=0$), we set a range with the lower and upper  bounds: $g_{ij}\in[g_{min},g_{max}]=[0.0001,1.0]$. Then, they are updated according to:
\begin{equation}\label{eq:8}
g_{ij} \to g_{ij} + \lambda\Delta g_{ij}(\Delta t_{ij}).
\end{equation}
In our simulations, we consider that the STDP update
rules are not applied to connections that are initially non-existent, i.e., we do not consider the creation of synapses. Moreover, if neurons are disconnected, they will remain disconnected throughout the total simulation time. The parameter $\lambda$ represents the learning rate. It was found that small learning rates led to more robust learning \cite{masquelier2008spike}. Hence, in this work, we choose a small learning rate (i.e., $\lambda=0.0001$) which, by the way, also simulate the effect of STDP on the long-term evolution of neural network \cite{ren2012hopf}.

The synaptic modification $\Delta g_{ij}(\Delta t_{ij})$ depends on the relative time difference  $\Delta t_{ij}=(t^{(post)}_i - t^{(pre)}_j)$ between the nearest-spike times of the post-synaptic neuron $i$ and the pre-synaptic neuron $j$. In this work, we consider STDP with an asymmetric Hebbian time window for the synaptic modification $\Delta g_{ij}(\Delta t_{ij})$ given by \cite{bi1998synaptic,kim2018stochastic}:
\begin{equation}\label{eq:10a}
 \Delta g_{ij}(\Delta t_{ij})= 
  \left\{
\begin{array}{ll}
\displaystyle{A_1e^{-\Delta t_{ij}/\tau_1},~\text{if}~\Delta t_{ij}>0 }\\[2.0mm]
\displaystyle{- A_2e^{\Delta t_{ij}/\tau_2},~\text{if}~\Delta t_{ij}<0}\\[2.0mm]
0, ~\text{if}~\Delta t_{ij}=0
\end{array} 
\right. , 
\end{equation}
where long term potentiation (LTP --- strengthening of synapses) occurs for $\Delta t_{ij}>0$ (i.e., a post-synaptic spike follows a pre-synaptic spike), long term depression (LTD --- weakening of synapses) occurs for $\Delta t_{ij}<0$ (i.e., a post-synaptic spike precedes a pre-synaptic spike), and no synaptic modifications for $\Delta t_{ij}=0$ (i.e., a post-synaptic spike coincides a pre-synaptic spike). The amount of synaptic modification (i.e., strengthening or weakening) is limited by the adjusting rate parameters $A_1$ and $A_2$. $\tau_1$ and $\tau_2$ determine the temporal window for synaptic modification.

\begin{figure}
\centering
\includegraphics[width=8.0cm,height=4.5cm]{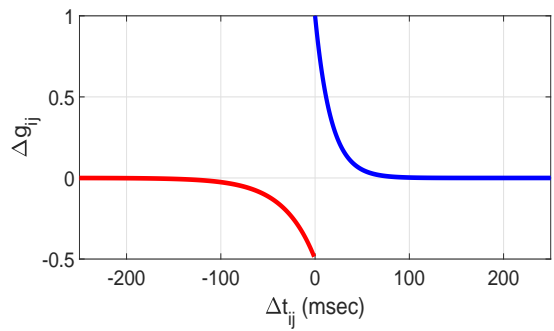}
\caption{
Plot of synaptic modification $\Delta g_{ij}$ versus $\Delta t_{ij}$.  The blue and red curves represent LTP and LTD, respectively (see main text). $A_1=1.0$, $A_2=0.5$, $\tau_1= 20.0$,  $\tau_2 = 20.0$.}

\label{fig:2}
\end{figure}

Fig.~\ref{fig:2} shows the asymmetric Hebbian time window for the synaptic modification $\Delta g_{ij}(\Delta t_{ij})$ given by Eq.\eqref{eq:10a}. We see in both cases that $\Delta g_{ij}$ varies, depending on the relative time difference $\Delta t_{ij}$ between the nearest spike times of the post-synaptic neuron $i$ and the pre-synaptic neuron $j$. Some experimental studies \cite{zhang1998critical,froemke2002spike,wolters2003temporally} have shown that $A_2\tau_2>A_1\tau_1$ ensures dominant depression of synapses, otherwise dominant potentiation. Throughout our studies, we fix the depression adjusting rate parameter at $A_2=0.5$ and the potentiation temporal window parameter at $\tau_1=20.0$ $\mathrm{msec}$.  Then, we choose the potentiation adjusting rate parameter $A_1\in[0.001,1.0]$ and the depression temporal window parameter $\tau_2\in[1.0,30.0]$ as the alterable parameters of the STDP learning rule. Thus, with the fixed parameter values of $A_2$ and $\tau_1$ and the inequality $A_2\tau_2>A_1\tau_1$ ($A_2\tau_2 \leq A_1\tau_1$), we can have a dominant LTD (LTP) over LTP (LTD) by changing the values of the alterable parameters $A_1$ and $\tau_2$ within the range of values chosen above.
%%%%%%%%%%%%%%%%%%%%%%%%%%%%%%%%%%%%%%%%%%%%%%%%%%%%%%%%%%%%%%%
\section{Computational methods}\label{Sec. III}
%%%%%%%%%%%%%%%%%%%%%%%%%%%%%%%%%%%%%%%%%%%%%%%%%%%%%%%%%%%%%%%
To measure the degree of regularity of the spiking activity induced by the mechanism of CR in the networks, we use the coefficient of variation --- an important statistical measure based on the time intervals between spikes \cite{pikovsky1997coherence, masoliver2017coherence}. It is related to the timing precision of information processing in neural systems \cite{pei1996noise}.  We numerically integrate the set of stochastic differential equations in Eq.\eqref{eq:1} with the Hebbian STDP rule of Eq.\eqref{eq:10a} by using the Runge-Kutta algorithm for stochastic processes \cite{,kasdin1995runge} and the Box-Muller algorithm \cite{knuth1973art}.
We set the integration time step to $dt = 0.01$ $\mathrm{msec}$ for a total time of $T=2.5\times10^{3}$ $\mathrm{msec}$. 
Averages are taken over 50 different realizations of the initial conditions. For each realization,  we choose random initial points $[V_i(0),x_i(0)]$ for the $i$th neuron with uniform probability in the range of $V_i(0)\in(-75, 40)$, 
$x_i(0)\in(0,1)$, $x_{i}(0)=\{m_{i}(0),h_{i}(0),n_{i}(0)\}$. The initial synaptic weights $g_{ij}(t=0)$, are normally distributed with mean and standard deviation equal to $0.1$ and $0.02$, respectively. The source code used for the simulations is provided in ref. \cite{stdp_code}.

We use the coefficient of variation ($\mathrm{CV}$) given by the normalized standard deviation of the mean interspike interval (ISI) \cite{pikovsky1997coherence}. For $N$ coupled neurons, $\mathrm{CV}$ is given by \cite{masoliver2017coherence}
\begin{equation}\label{eq:15a}
    \mathrm{CV} =
    \frac{\sqrt{\overline{\langle \mathrm{\tau}^2 \rangle} - \overline{\langle \mathrm{\tau} \rangle^2}}}
        {\overline{\langle \mathrm{\tau} \rangle}},
\end{equation}
where
$\overline{\langle \mathrm{\tau} \rangle} = N^{-1} \sum_{i=1}^N \langle \mathrm{\tau}_i \rangle$ and 
$\overline{\langle \mathrm{\tau}^2 \rangle} = N^{-1}\sum_{i=1}^N \langle \mathrm{\tau}_i^2 \rangle$, with
$\langle \mathrm{\tau}_i \rangle$ 
and 
$\langle \mathrm{\tau}_i^2 \rangle$ representing
the mean and mean squared ISI (over time),  $\mathrm{\tau}_i = t_i^{\ell+1}-t_i^{\ell}>0$, of neuron $i$.
We determine the spike occurrence times from the instant the when the average membrane potential $V_{ave}$ crosses the threshold $V_{ave}=20.0$  $\mathrm{mV}$.  
The $\mathrm{CV}$ will be the higher the more variable the mean ISIs are. Thus, since Poisson spike train events are independent and all have a normalized standard deviation of unity (i.e., $\mathrm{CV}=1$), they can be used as reference for the average variability of spike trains of the network \cite{gabbiani1998principles}. 
When $\mathrm{CV}>1$, the average variability of spike trains of the network is more variable than a Poisson process. When $\mathrm{CV} < 1$, the average spiking activity of the network becomes more coherent, with $\mathrm{CV}=0$ corresponding to perfectly periodic spike trains.

The degree of synchronization of the spiking activity of the membrane potentials in the neural network would be quantified by the standard deviation $\Omega$, which is defined as \cite{gong2006ordering}:
\begin{equation}\label{eqn:12a}
\begin{split}
\left\{\begin{array}{lcl}
\Omega = \big[\langle\rho(t)\rangle\big],\\[2.0mm]
\rho(t) = \displaystyle{\sqrt{\frac{\Bigg[\displaystyle{\frac{1}{N}\sum\limits_{i=1}^N\Big(V_i(t)\Big)^2-\Bigg(\frac{1}{N}\sum\limits_{i=1}^NV_i(t)\Bigg)^2}\Bigg]}{N-1}}},
\end{array}\right.
\end{split}
\end{equation}
where $\rho(t)$ measures the synchronization of the membrane potential variable $V_i$ at a given time $t$. The angle brackets $\langle \cdot \rangle$ denotes the average over time, and the square brackets $[\cdot]$ the average over different realizations of the networks for each set of parameter values.  The value of $\Omega$ is an excellent indicator for numerically measuring the spatiotemporal synchronization of excitations, hence revealing different synchronization levels and related transitions in the network. 
From Eq.\eqref{eqn:12a}, smaller values of $\Omega$ indicate higher degrees of synchronization. 
Accordingly, when $\Omega=0$, the network reaches complete synchronization. 

It is worth pointing out that the synchronization of the actual values of the membrane potential is more
general than the synchronization of the spiking times of the membrane potential. That is, if in the
network, the actual values of the membrane potential of all neurons are the same at all times
(i.e., membrane potential synchronization), then the spiking times of all neurons (which is
defined by the times of the crossing of the actual values of the mean membrane potential passing the predefined threshold $V_{ave}=20.0$ $\mathrm{mV}$) must also be the same at all times (i.e., spiking time synchronization).
Whereas, if the network synchronizes only the spiking times of the neurons, this does not
necessarily mean that the neurons have the same actual values of the membrane potential (i.e.,
the amplitudes of the membrane potentials might remain uncorrelated while the spiking times
are synchronized). More importantly, we choose membrane potential synchronization defined in Eq.\eqref{eqn:12a} over spiking time
synchronization because it would enable us to directly compare our results with those in \cite{gong2005optimal}
which also uses $\Omega$ to measure synchronization.

In our simulations, we use small-world network generated by Watts-Strogatz algorithm \cite{watts1998collective,strogatz2001exploring}. We consider an inhibitory directed Watts-Strogatz small-world network (SWN), composed of $N=100$ HH interneurons equidistantly placed on a one-dimensional ring of radius $N/2\pi$. We note that the SWN interpolates between a regular lattice with high clustering (corresponding to the case of $\beta = 0$) and a random graph with short average path length (corresponding to the case of $\beta = 1$) via random uniform rewiring with the probability $\beta$. We note that the data for calculating the $\mathrm{CV}$ and $\Omega$ are recorded only after a sufficiently long transient time (i.e., $T\geq 7.5\times10^{2}$ $\mathrm{msec}$).

%%%%%%%%%%%%%%%%%%%%%%%%%%%%%%%%%%%%%%%%%%%%%%%%%%%%%%%%%%%%%%%
\section{Results and discussion}\label{Sec. IV}
\subsection{Interplay with network parameters}
%%%%%%%%%%%%%%%%%%%%%%%%%%%%%%%%%%%%%%%%%%%%%%%%%%%%%%%%%%%%%%%
In this subsection, we present the numerical results on the interplay between CR and SS in terms of the network parameters, i.e., the average degree $\langle k \rangle$ and the rewiring probability $\beta$, when the STDP learning rule is dominant potentiation, i.e., $A_2\tau_2 < A_1\tau_1$. Because we want to simultaneously optimize the degree of CR and SS, we have chosen $A_2$, $\tau_2$, $A_1$, and $\tau_1$ such that we have potentiation which, as we shall see later, (i) always enhances the degree of SS, and (ii) depending on the membrane patch area $S$, also enhances the degree of CR. 
%%%%%%%%%%%%%%%%%%%%%%%%%%%%%%%%%%%%%%%%%%%%%%%%%%%%%%%%%%%%%%%
\subsubsection{Average degree $\langle k \rangle$}
\begin{figure*}
\centering
\includegraphics[width=7.0cm,height=4.0cm]{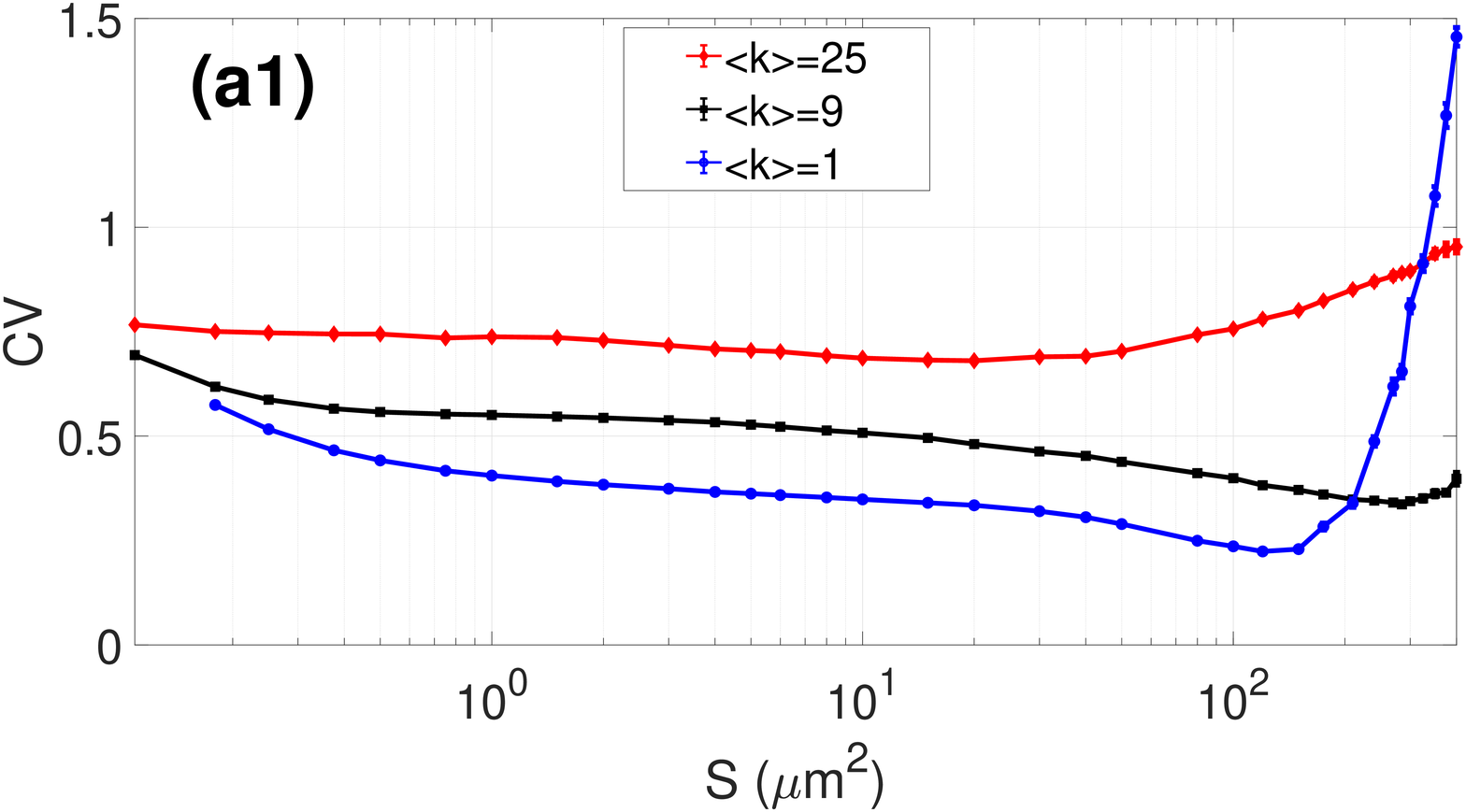}\includegraphics[width=7.0cm,height=4.0cm]{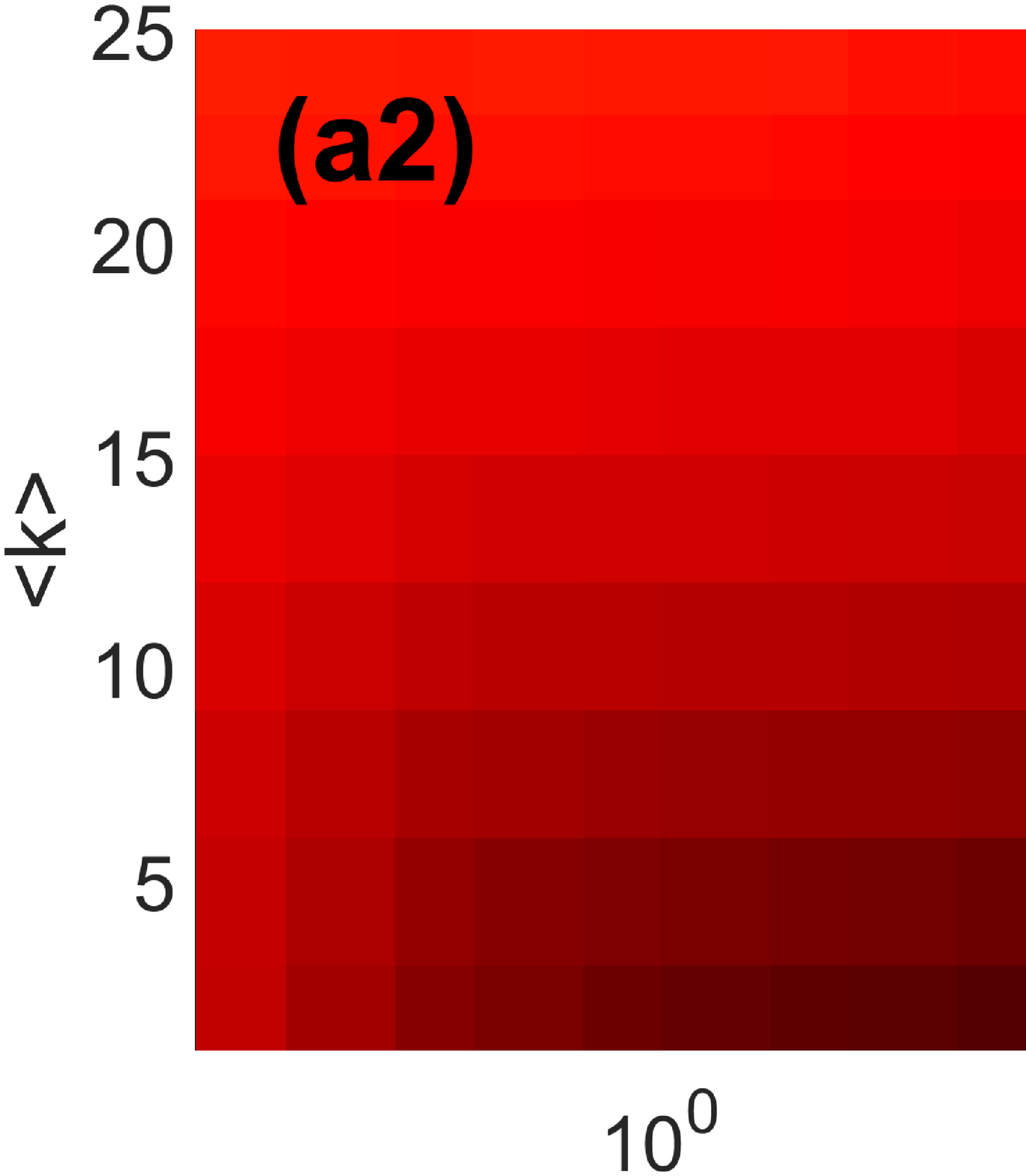}
\includegraphics[width=7.0cm,height=4.0cm]{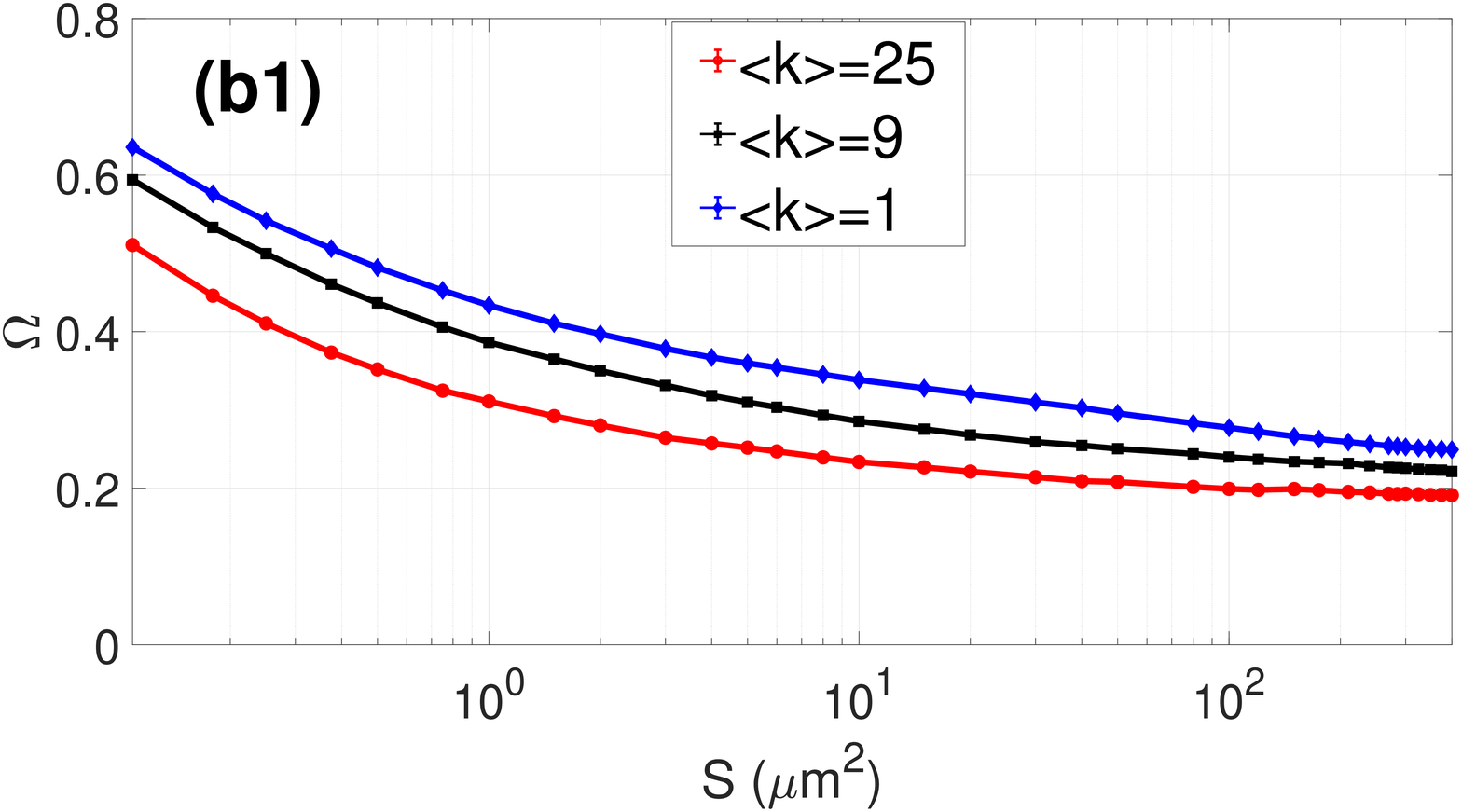}\includegraphics[width=7.0cm,height=4.0cm]{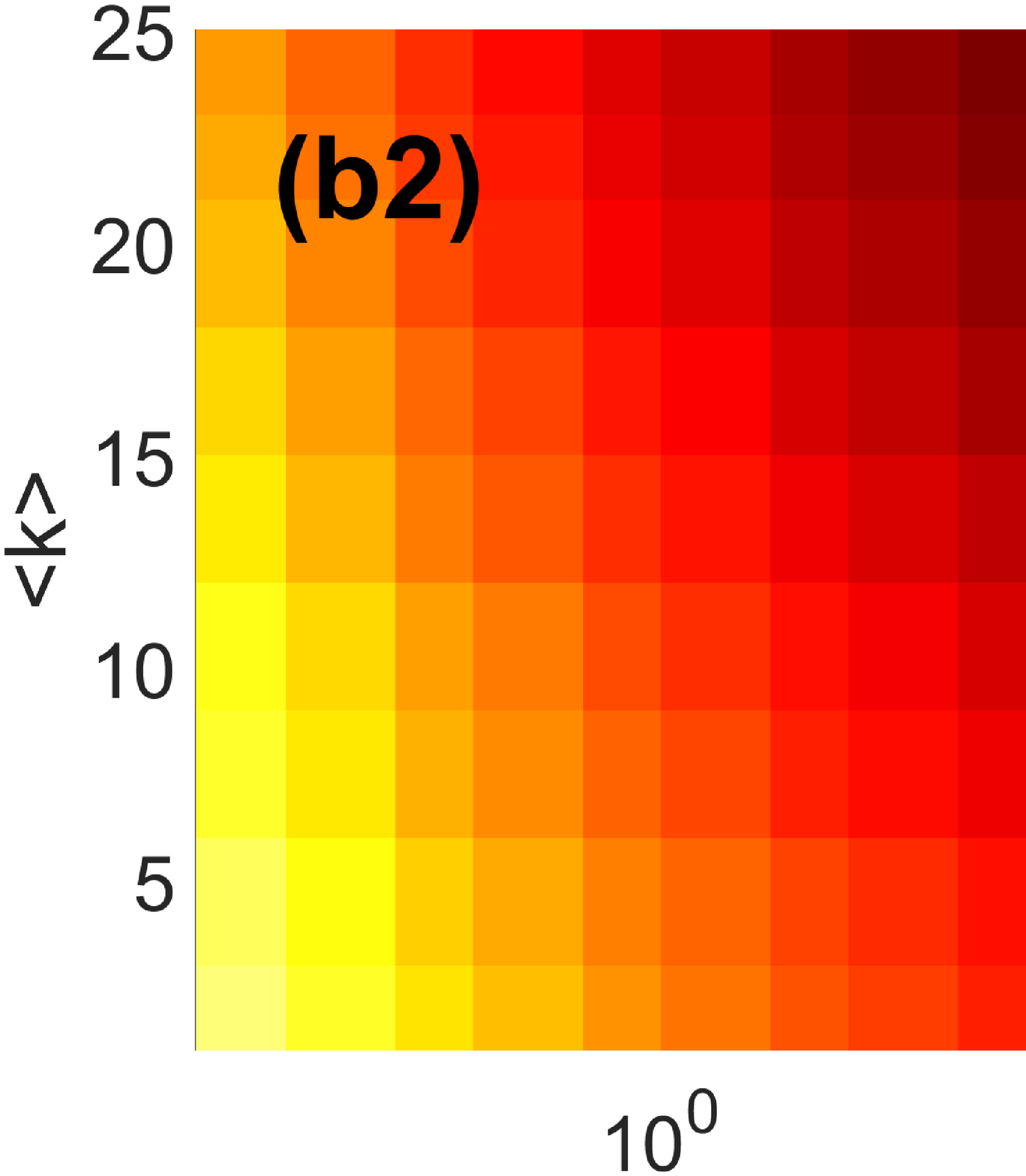}
\caption{
Variation of the coefficient of variation $\mathrm{CV}$ and the degree of stochastic synchronization $\Omega$ w.r.t. the membrane patch area $S$ and the average degree $\langle k \rangle$. Panels \textbf{(a1)} and \textbf{(a2)} show a non-monotonic behavior of the $\mathrm{CV}$ curves w.r.t. $S$.
Panels \textbf{(b1)} and \textbf{(b2)} show a monotonic behavior of the $\Omega$ curves w.r.t. $S$. $\beta=0.25$, $A_1=1.0$, $\tau_1= 20.0$, $A_2=0.5$, $\tau_2 = 20.0$.}

\label{fig:3}
\end{figure*}

%%%%%%%%%%%%%%%%%%%%%%%%%%%%%%%%%%%%%%%%%%%%%%%%%%%%%%%%%%%%%%%
Fig.~\ref{fig:3} shows the variation of $\mathrm{CV}$ and $\Omega$ with the average degree parameter $\langle k \rangle$ and the membrane patch area $S$ which is on the logarithmic scale in all the results presented in this paper. It is worth keeping in mind that the membrane patch area $S$ is, according to Eq.\eqref{eq:3}, inversely proportional to the strength of the channel noise, i.e., larger (smaller) $S$ represent weaker (stronger) channel noise intensities.

In Figs.~\ref{fig:3}\textbf{(a1)} and \textbf{(a2)}, the $\mathrm{CV}$ shows non-monotonic behaviors --- characteristic of CR --- as $S$ increases (i.e., as the noise intensity decreases). First, we observe that larger values of $\langle k \rangle$ shift the $\mathrm{CV}$ curves to higher values, thus deteriorating the degree of CR. This indicates that the sparser the network is, the less the neurons are affected by the random fluctuations of their neighbors, and hence the more regular is their spiking activity. 

Secondly, we observe that as $\langle k \rangle$ increases in the interval $[1,15]$, the minima of the $\mathrm{CV}$ curves are shifted to the right. See, e.g., the blue (with $\langle k \rangle=1$) and the black (with $\langle k \rangle=9$) curves in Fig.~\ref{fig:3}\textbf{(a1)}. One observes that the minimum of the $\mathrm{CV}$ curve is shifted to the right as $\langle k \rangle$ increases. From the right-bottom corner of Fig.~\ref{fig:3}\textbf{(a2)}, we can also observe that this behavior (i.e., the shift of the minimum of the $\mathrm{CV}$ curves to the right) persist as $\langle k \rangle$ increases in the interval $[1,15]$. But as soon as $\langle k \rangle \geq15$, the minimum of the $\mathrm{CV}$ curves are shifted to back to the left, see, e.g., the minimum of the red curve  (with $\langle k \rangle=25$) in Fig.~\ref{fig:3}\textbf{(a1)}. This indicates that at larger $\langle k \rangle \geq15$, smaller $S$ (i.e., stronger noise intensities) slightly improve the degree of CR.

In Figs.~\ref{fig:3}\textbf{(b1)} and \textbf{(b2)}, the degree of SS is indicated by the $\Omega$ curves (for different values of $\langle k \rangle$) as $S$ changes. First, we observe that as $S$ increases, the degree of SS increases (as indicated by decreasing value of $\Omega$). One will, of course, expect an enhancement of the degree of synchronization of the spiking activity with decreasing noise intensity, provided that these weaker noise intensities are still strong enough to induce spiking from the excitable regime of the neurons in the network. It is worth noting that the synchronization of the spiking activity in our network is purely noise-induced --- stochastic synchronization.  We recall that the external current parameter is fixed at $I^e=6.0$ $\mathrm{\mu A/cm^2}$, a value which is below the subcritical Hopf bifurcation threshold $I^e_H=6.27$ $\mathrm{\mu A/cm^2}$. Hence, no oscillations can occur in the zero-noise network, i.e., the deterministic network is in a homogeneous quiescent state. Now, in the presence of noise, the spiking activities (from the homogeneous quiescent state) of the neurons are purely noise-induced, and consequently, the synchronization of their spiking activities is also purely noise-induced. In the absence of noise (or in the presence of an extremely weak noise intensity), no neuron in the network would be able to spike, and we get $\Omega=0$ (or $\Omega\approx0$), indicating a complete synchronization. This would, however, be a complete synchronization of the homogeneous quiescent state and not a complete synchronization of stochastic oscillations (which are non-existent). In this work, we are interested in stochastic synchronization. With the largest value of $S=400$ (i.e., the weakest noise intensity) considered in our simulations, the network always produces, on average, at least 4 spikes. This value of $S$ permits us to avoid quiescent synchronization in the network and to consider only the (stochastic) synchronization of the noise-induced spiking activity.

\begin{figure}
\centering
\includegraphics[width=7.0cm,height=4.0cm]{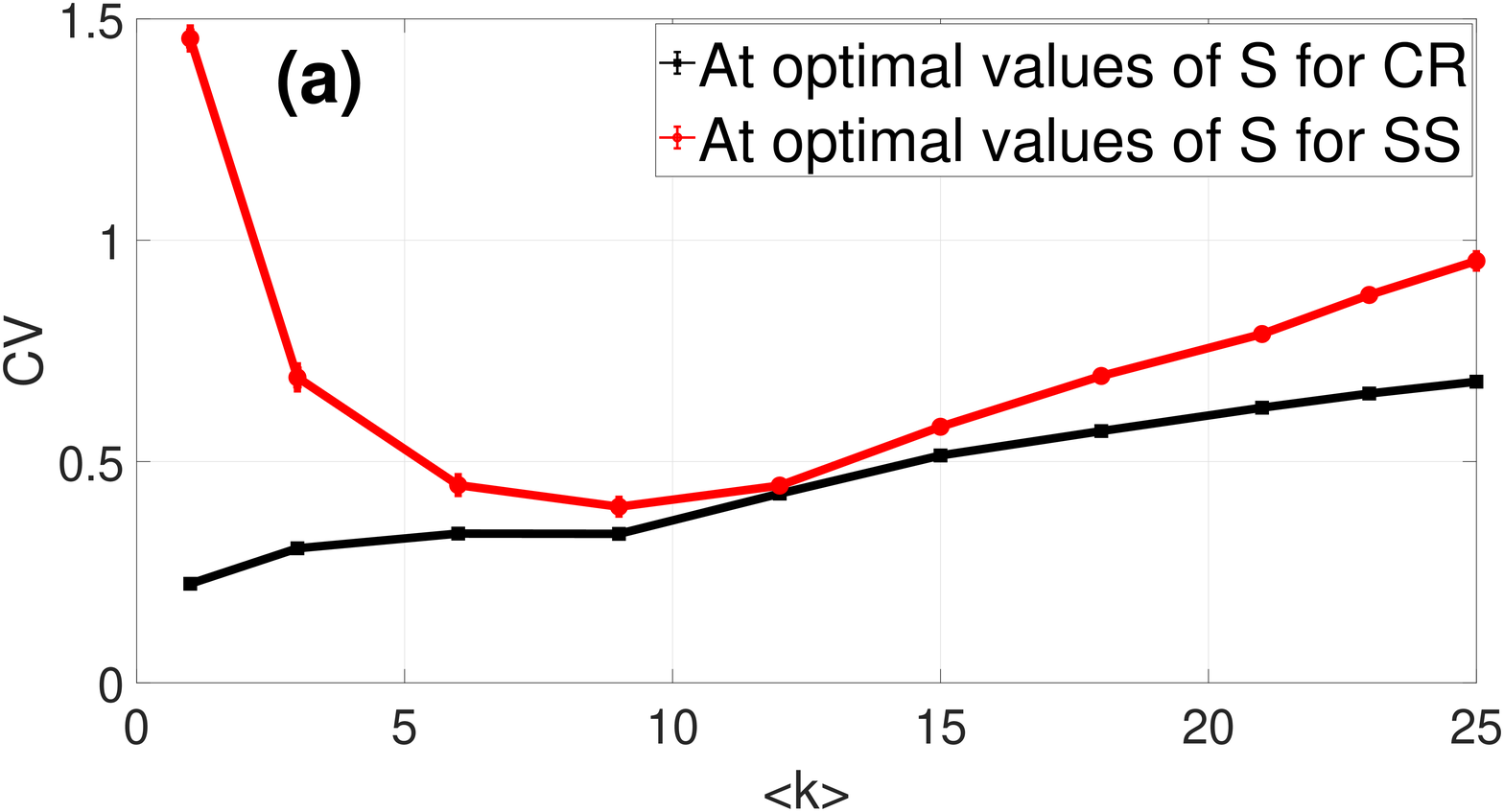}
\includegraphics[width=7.0cm,height=4.0cm]{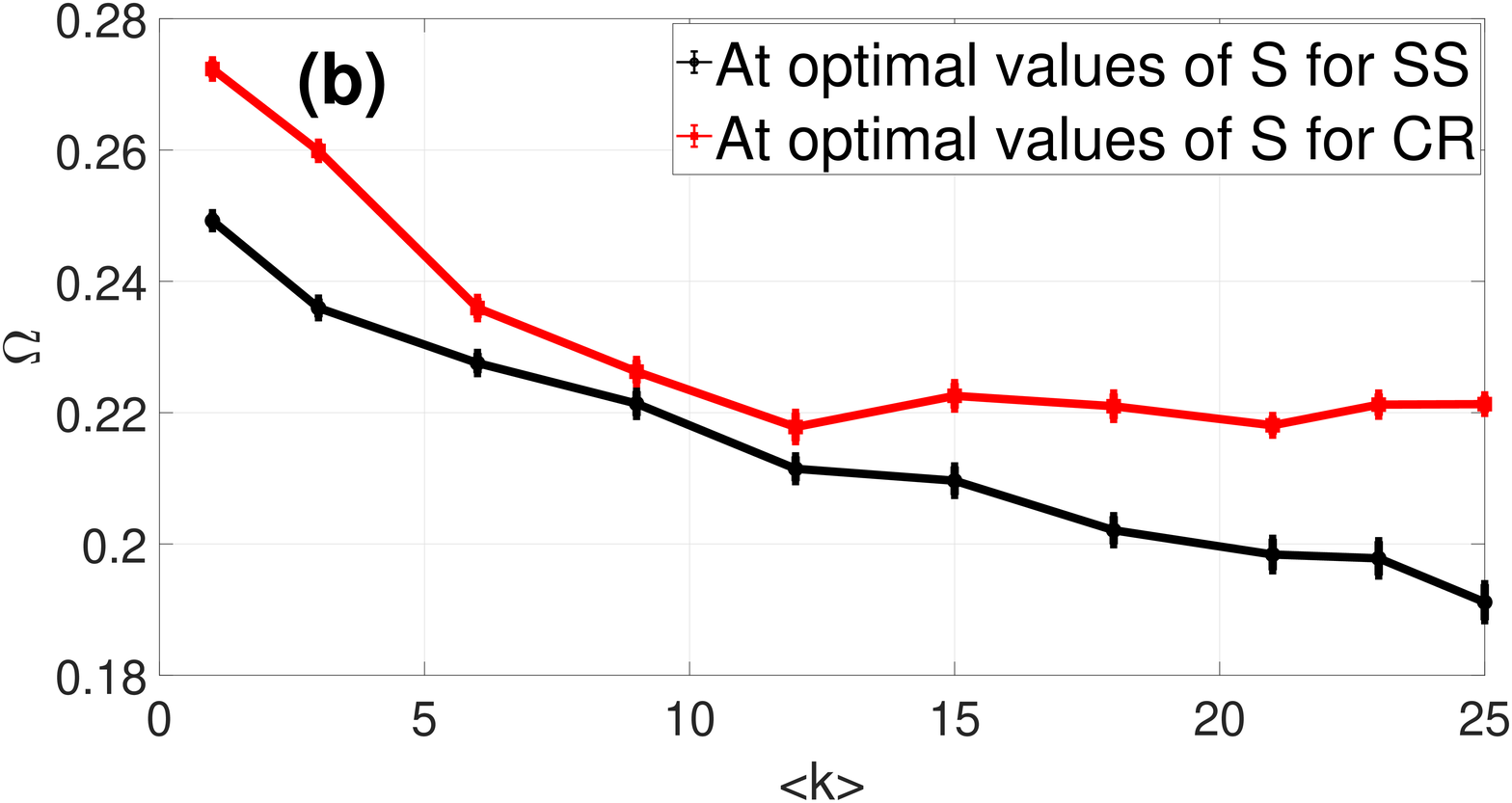}
\caption{Variation of the coefficient of variation $\mathrm{CV}$ (panel \textbf{(a)}) and the degree of stochastic synchronization $\Omega$ (panel \textbf{(b)}) w.r.t. the average degree $\langle k \rangle$ at their optimal and non-optimal membrane patch area $S$. $\beta=0.25$, $A_1=1.0$, $\tau_1= 20.0$, $A_2=0.5$, $\tau_2 = 20.0$.}

\label{fig:4}
\end{figure}

 Larger values of $\langle k \rangle$ mean that more neurons can interact with each other, and hence their ability to synchronize their spiking activity is enhanced. We observe in Figs.~\ref{fig:3}\textbf{(b1)} and \textbf{(b2)} (in contrast to CR in Figs.~\ref{fig:3}\textbf{(a1)} and \textbf{(a2)}) that larger values of $\langle k \rangle$ induce a higher degree of SS because $\Omega$ becomes smaller. Furthermore, we observe,
 by comparing the $\mathrm{CV}$ and $\Omega$ curves in Fig.~\ref{fig:3}, that the best degree of SS occurs at the largest value of $S=400$. While the best degree of CR (i.e., minimum of the $\mathrm{CV}$) occurs at intermediate values of $S$, which shall hence forth be referred to as the resonant values of the membrane patch area. 
With this in mind, we now investigate how to use the average degree $\langle k \rangle$ to enhance both CR and SS at values of $S$ where each of these phenomena is not optimal. 
We start with improving the degree of CR at $S=400$ (i.e., the value of $S$ at which the degree of SS is the best). The black curve in Fig.~\ref{fig:4}\textbf{(a)} shows the variation of $\mathrm{CV}$ with $\langle k \rangle$ at the resonant values of  $S$ [i.e., the resonant values of $S$ at which the degree of CR is the best]. We notice a monotonic increase of this black curve, meaning that the  optimal CR deteriorates as $\langle k \rangle$ increases. The minimum of the red curve in Fig.~\ref{fig:4}\textbf{(a)} indicates values of $\langle k \rangle$ for which the best degrees of SS and CR occur simultaneously at the largest value of $S(=400)$. The non-monotonic shape of the red curve indicates that at $S=400$, increasing $\langle k \rangle$ does not necessarily deteriorate the CR.  In fact, it shows that at this value of  $S$, we can enhance the degree of the CR by increasing $\langle k \rangle$ (which would always enhance SS) up to a certain intermediate value. Therefore, we can simultaneously achieve high degrees of CR and SS at the largest $S$, when the average degree of the network is at some intermediate value. In our network, at $S=400$, the best degree of CR occurs at $\langle k \rangle=9$, see the red curve in Fig.~\ref{fig:4}\textbf{(a)}.  

Now we investigate how to use $\langle k \rangle$ to improve the degree of SS at resonant values of $S$, each of which represents the value at which the degree of CR is best, at the corresponding values of $\langle k \rangle$.
In Fig.~\ref{fig:4}\textbf{(b)}, the black curve shows the variation of degree of SS at the resonant values of $S$ as $\langle k \rangle$ increases. It is clear that larger $\langle k \rangle$ improves SS. The fact that the red curve in Fig.~\ref{fig:4}\textbf{(b)} decreases monotonically as $\langle k \rangle$ increases, indicates that we can also simultaneously achieve high degrees of both CR and SS at intermediate values of $S$.

%%%%%%%%%%%%%%%%%%%%%%%%%%%%%%%%%%%%%%%%%%%%%%%%%%%%%%%%%%%%%%%
\subsubsection{Rewiring probability $\beta$}
\begin{figure*}
\centering
\includegraphics[width=7.0cm,height=4.0cm]{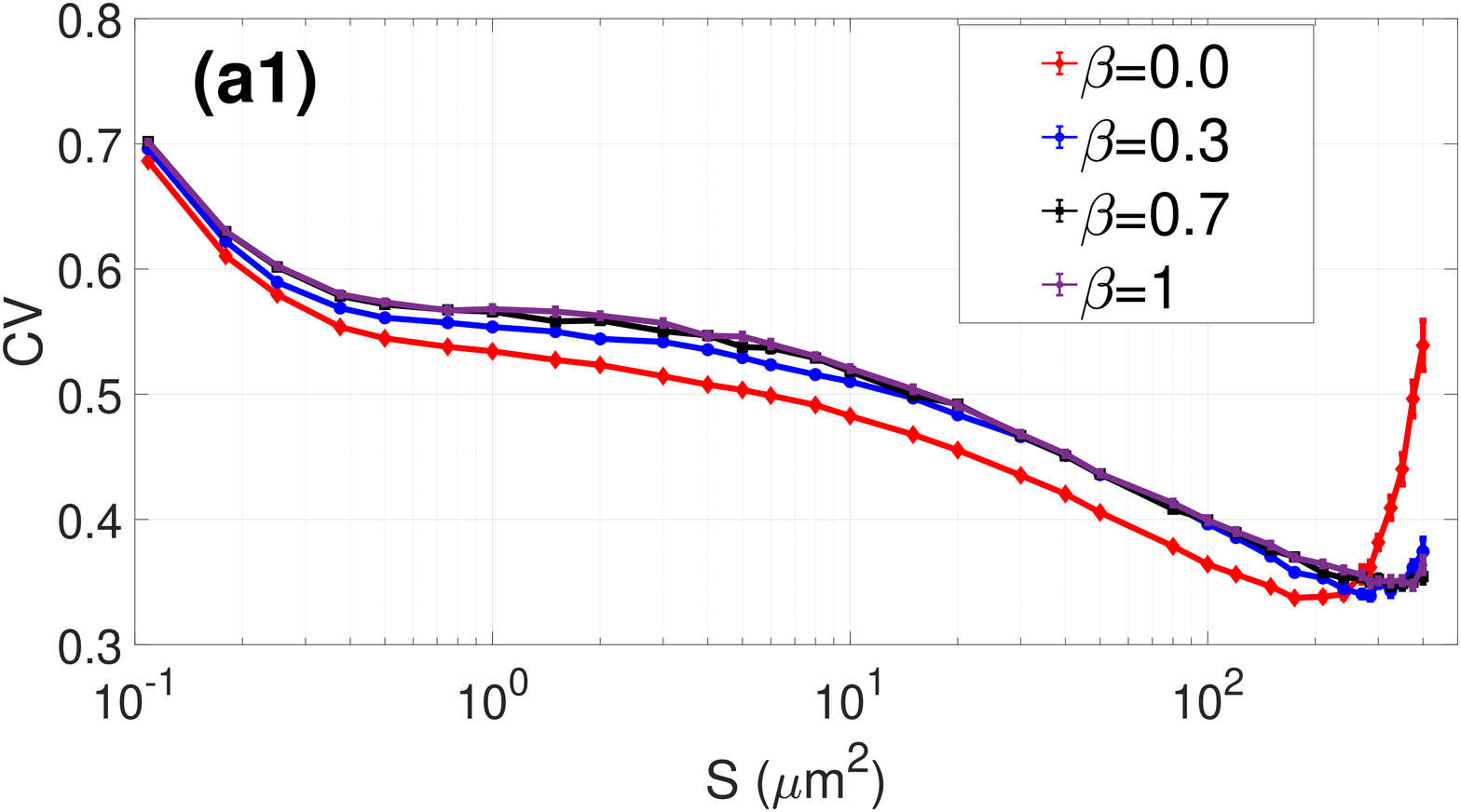}\includegraphics[width=7.0cm,height=4.0cm]{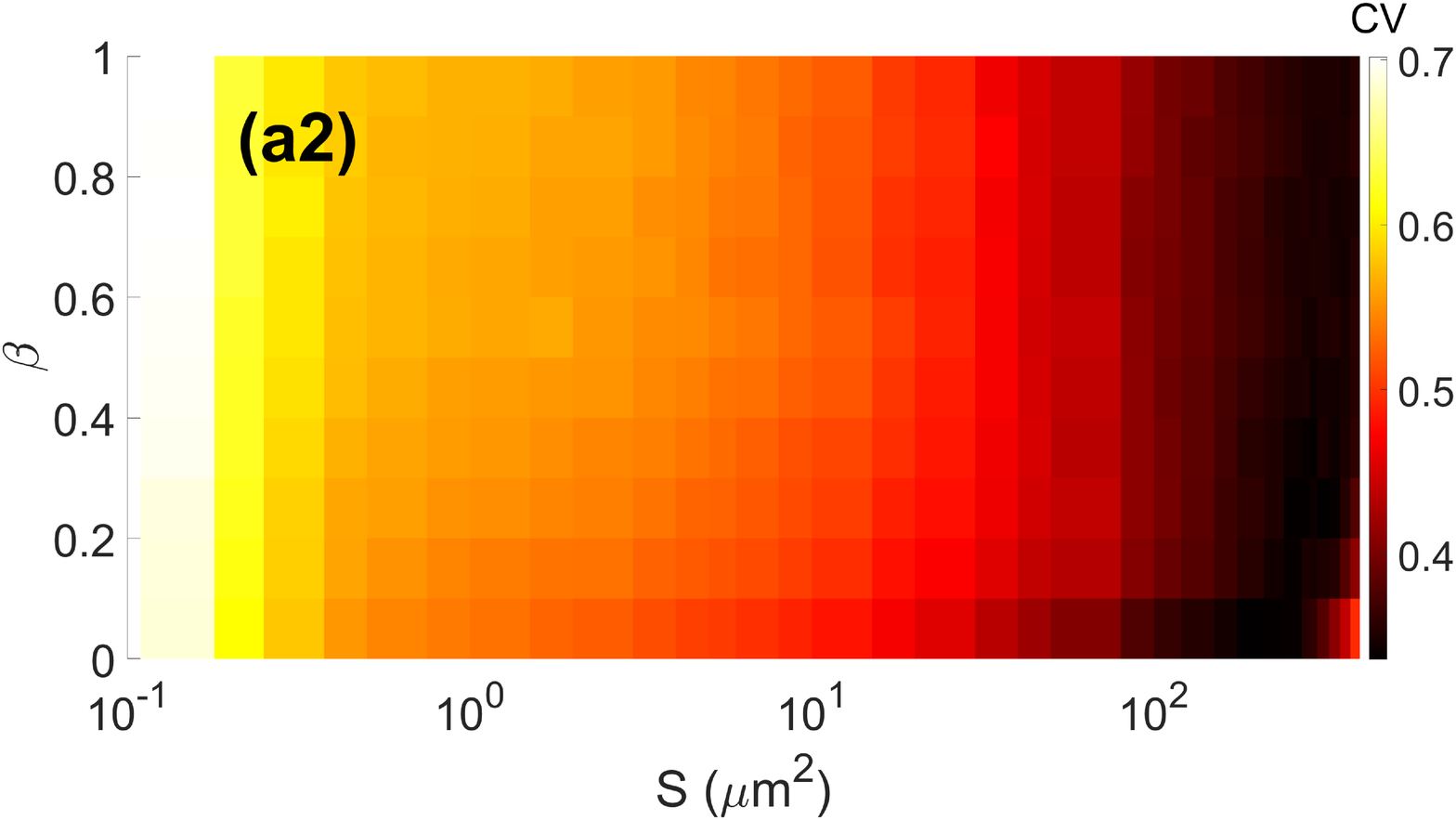}
\includegraphics[width=7.0cm,height=4.0cm]{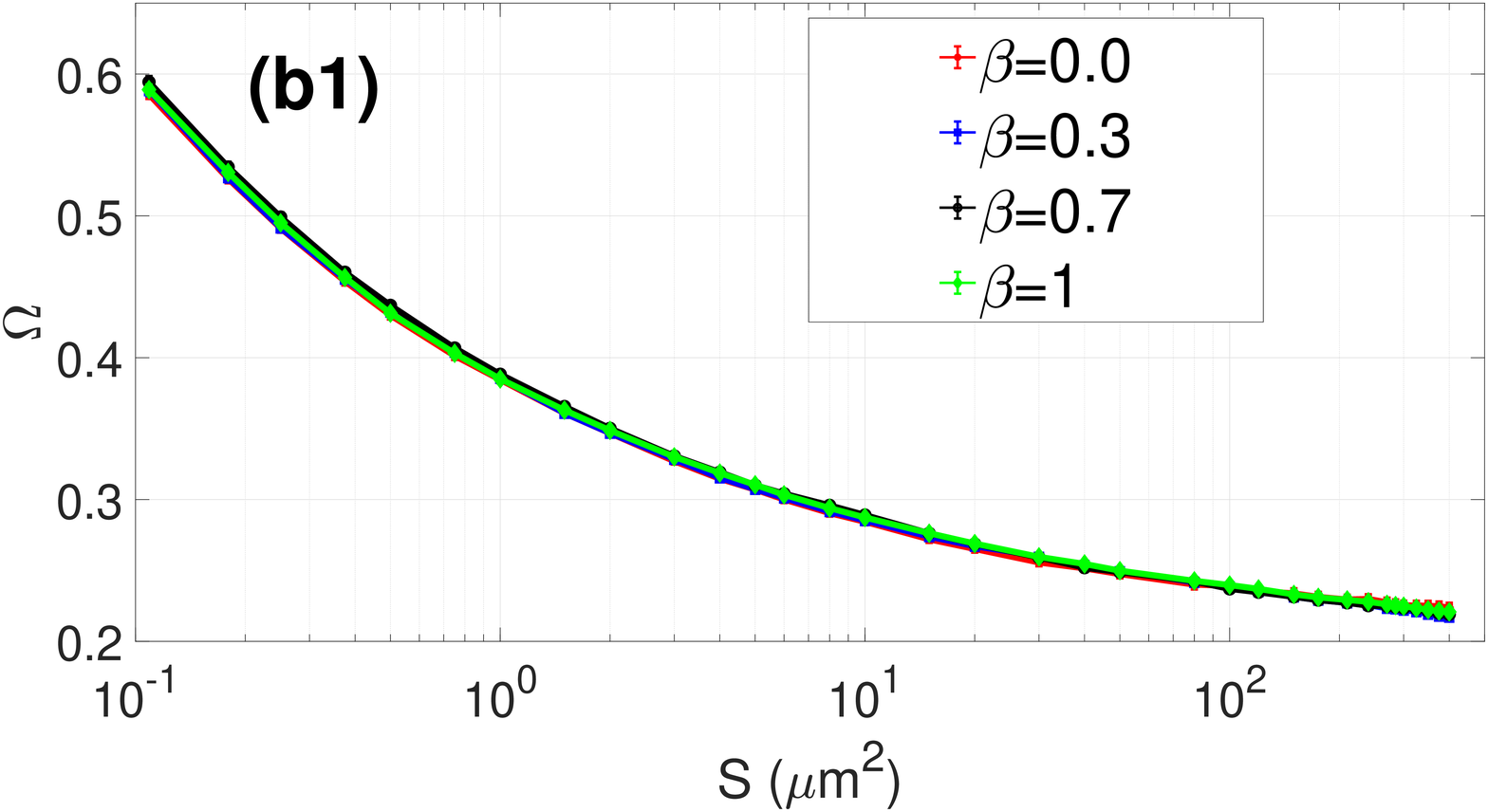}\includegraphics[width=7.0cm,height=4.0cm]{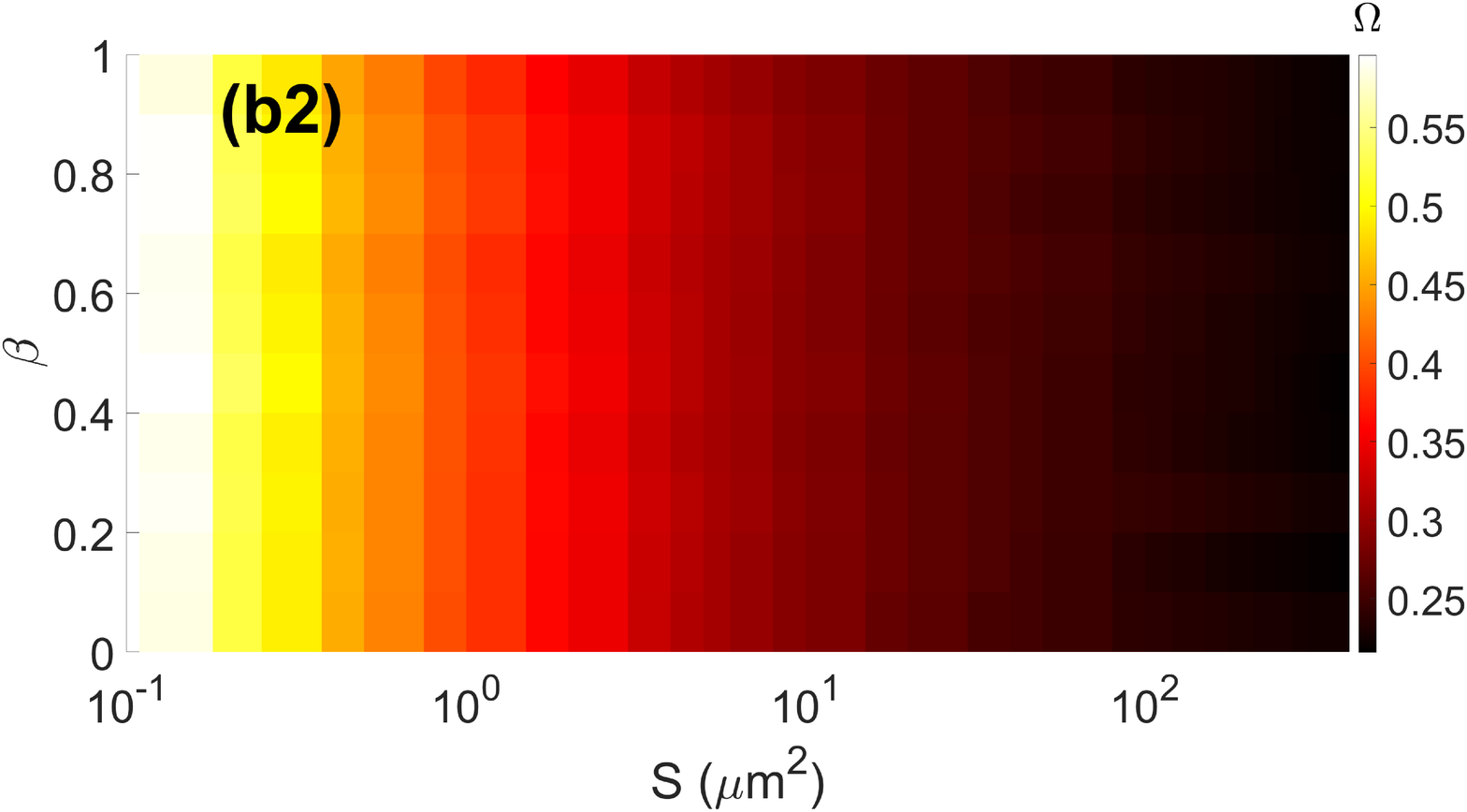}
\caption{
Variation of $\mathrm{CV}$ and $\Omega$ w.r.t. $S$ and the rewiring probability $\beta$. In panels \textbf{(a1)} and \textbf{(a2)}, as $\beta$ increases, the degree of CR at the optimal values of $S$ slightly changes. In panels \textbf{(b1)} and \textbf{(b2)}, larger $S$ increases the degree of SS, and as $\beta$ increases, the degree of SS is essentially the same. $\langle k \rangle=9$, $A_1=1.0$, $\tau_1= 20.0$, $A_2=0.5$, $\tau_2= 20.0$.}

\label{fig:5}
\end{figure*}

%%%%%%%%%%%%%%%%%%%%%%%%%%%%%%%%%%%%%%%%%%%%%%%%%%%%%%%%%%%%%%%
Fig.~\ref{fig:5} shows the variation of $\mathrm{CV}$ and $\Omega$ with the rewiring probability parameter $\beta$ and the membrane patch area $S$. Again since we want to simultaneously optimize the degree of CR and SS, we choose the value of $\langle k \rangle=9$ for which the red curve in Fig.~\ref{fig:4}\textbf{(a)} is minimum.

In Figs.~\ref{fig:5}\textbf{(a1)} and \textbf{(a2)}, the $\mathrm{CV}$ curves (for different values of $\beta$) show a non-monotonic behavior --- characteristic of CR --- as $S$ increases. The simulations show that the more random the network becomes (i.e., as $\beta$ increases), the lower the degree of CR becomes, as indicated by the higher values of the $\mathrm{CV}$ curves. However, this decrease in the degree of the coherence as $\beta$ increases is not very pronounced. 
Furthermore, when the network is regular (i.e., when $\beta=0$), the highest degree of CR occurs at a slightly smaller value of $S$ than in the cases of a small-world network (i.e., when $0<\beta<1$) and completely random network (i.e., when $\beta=1$).

Figs.~\ref{fig:5}\textbf{(b1)} and \textbf{(b2)} show the variation in the degree of SS at different values of $\beta$ as $S$ increases. Higher values of $S$ increase the degree of SS, but higher values of $\beta$ do not essentially change the degree of SS. In particular, at $S=400$, $\Omega$ fluctuates within a very thin interval $[0.2157,0.2239]$ as $\beta$ increases. From Fig.~\ref{fig:5}, it can be seen that as $\beta$ increases, the best degree of CR occurs at some intermediate values of $S$, whereas the best degree of SS occurs at the largest value of $S=400$, irrespective of the value of $\beta$. With this in mind, we now investigate how to use the rewiring probability $\beta$ to enhance both CR and SS at the values of $S$ at which each of these phenomena are not optimal.

\begin{figure}
\centering
\includegraphics[width=7.0cm,height=4.0cm]{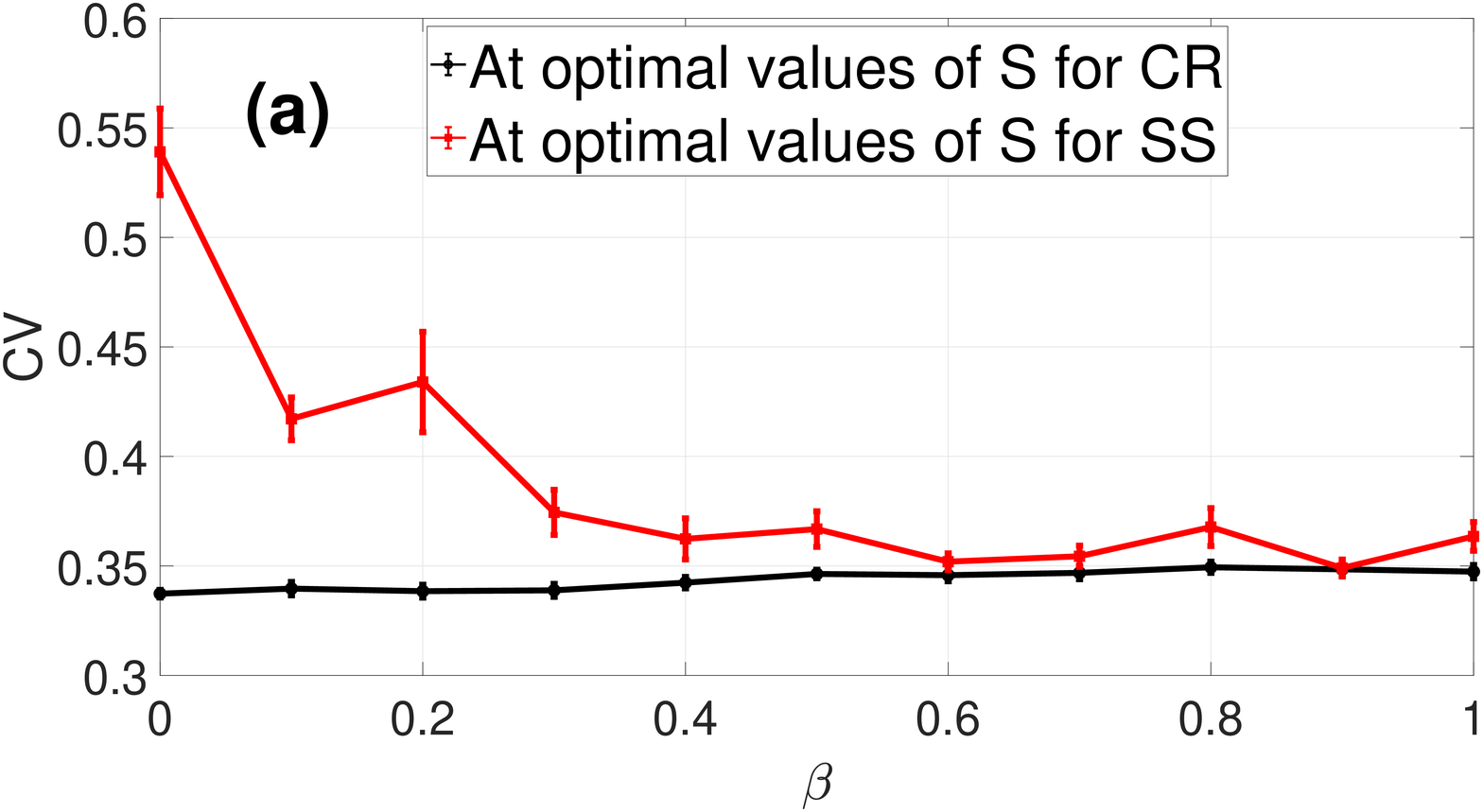}
\includegraphics[width=7.0cm,height=4.0cm]{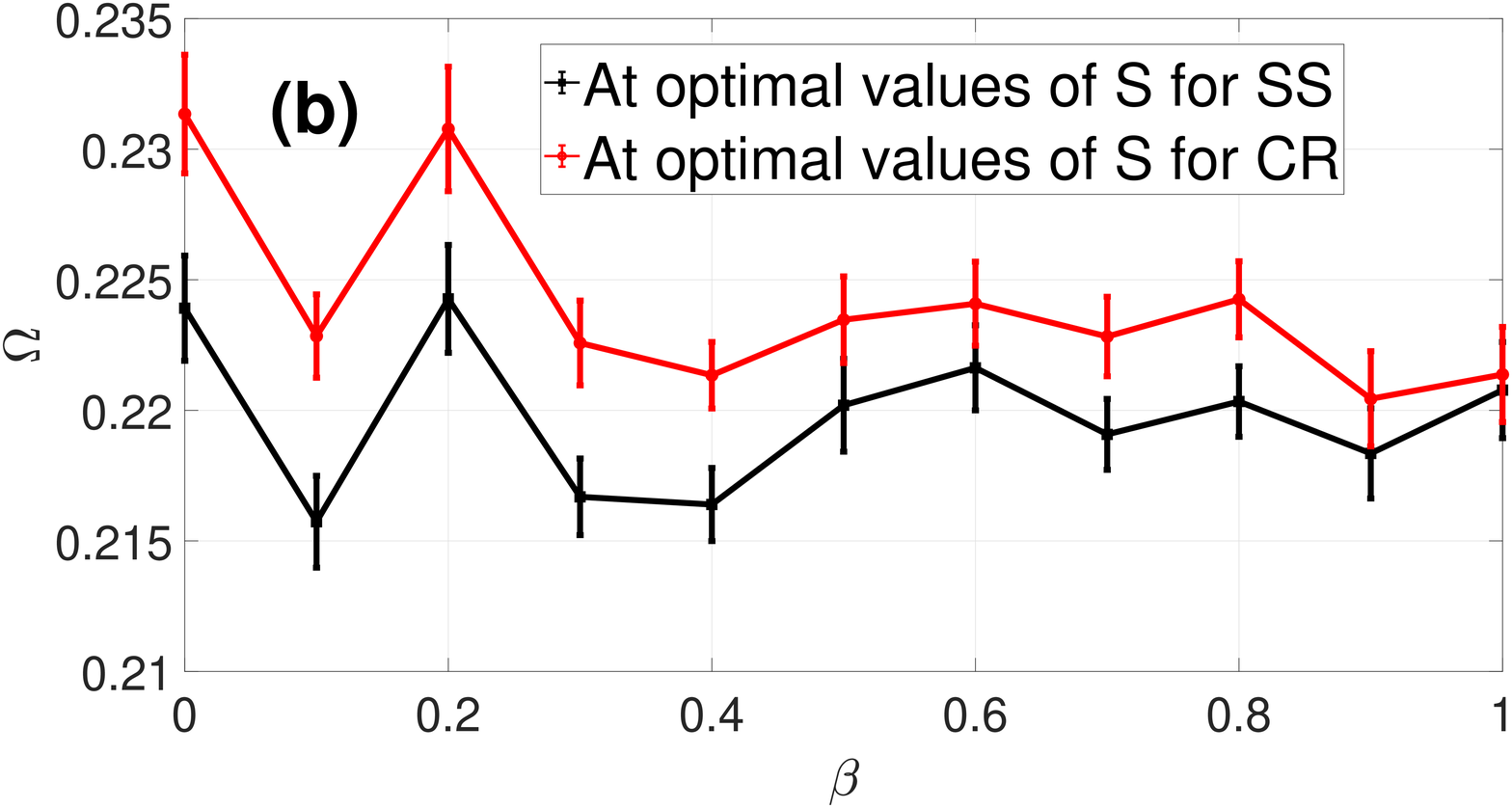}
\caption{
Variation of the coefficient of variation $\mathrm{CV}$ (panel \textbf{(a)}) and the degree of stochastic synchronization $\Omega$ (panel \textbf{(b)}) w.r.t. the rewiring probability $\beta$ at their optimal and non-optimal membrane patch area $S$. $\langle k \rangle=9$, $A_1=1.0$, $\tau_1= 20.0$, $A_2=0.5$, $\tau_2= 20.0$.}

\label{fig:6}
\end{figure}

In Fig.~\ref{fig:6}\textbf{(a)}, the black curve shows that increasing $\beta$ only slightly decreases the degree of CR at intermediate values of $S$. While the red curve indicates that increasing $\beta$ can increase the degree of CR at the largest value of $S$. And since the degree of SS is not very much affected by changes in $\beta$ (see, e.g., Fig.~\ref{fig:5}\textbf{(b1)}), we can therefore enhance the degree of CR at the largest value of $S$ by increasing $\beta$.

Fig.~\ref{fig:6}\textbf{(b)} indicates that the high degree of SS is essentially not affected by $\beta$ for all values of $S$. This can been seen from the red and black curves which are constant within error bars. Since (i) the degree of CR at the resonant values of $S$ is only slightly decreasing as $\beta$ increases (see black curve in Fig.~\ref{fig:6}\textbf{(a)}), (ii) the degree of CR at the largest value of $S$ increases as $\beta$ increases up to $0.6$, after which it stays essentially constant (see red curve in Fig.~\ref{fig:6}\textbf{(a)}), and (iii) the degree of SS fluctuates within a thin interval of relatively low values of $\Omega$ as $\beta$ increases, we can, therefore, maximize the enhancement of both CR and SS by choosing an arbitrary high value of $\beta$. Thus, in the sequel the rewiring probability is fixed at $\beta=0.9$.

 At this point, from Figs.~\ref{fig:6}\textbf{(b)}, it is worth pointing out that in comparison to the work in \cite{gong2005optimal}, where the HH neural network has no STDP, the synchronization indicator $\Omega$ shows a more significant dependence on the number of added random shortcuts than in the current work. This indicates that STDP stabilizes the effect of adding more random shortcuts (i.e., increasing the value of $\beta$) in the network on the degree of synchronization.

%%%%%%%%%%%%%%%%%%%%%%%%%%%%%%%%%%%%%%%%%%%%%%%%%%%%%%%%%%%%%%%
\subsection{Interplay with STDP parameters}
%%%%%%%%%%%%%%%%%%%%%%%%%%%%%%%%%%%%%%%%%%%%%%%%%%%%%%%%%%%%%%%
In this subsection, we present the numerical results on the interplay between CR and SS in terms of the STDP parameters, i.e., the potentiation adjusting rate parameter $A_1$, and the depression temporal window parameter $\tau_2$. Here, the network parameters are fixed at values at which both CR and SS are optimal, i.e., $\langle k \rangle=9$ and $\beta=0.9$.
%%%%%%%%%%%%%%%%%%%%%%%%%%%%%%%%%%%%%%%%%%%%%%%%%%%%%%%%%%%%%%%
\subsubsection{Potentiation adjusting rate $A_1$}
\begin{figure}
\centering
\includegraphics[width=8.0cm,height=4.0cm]{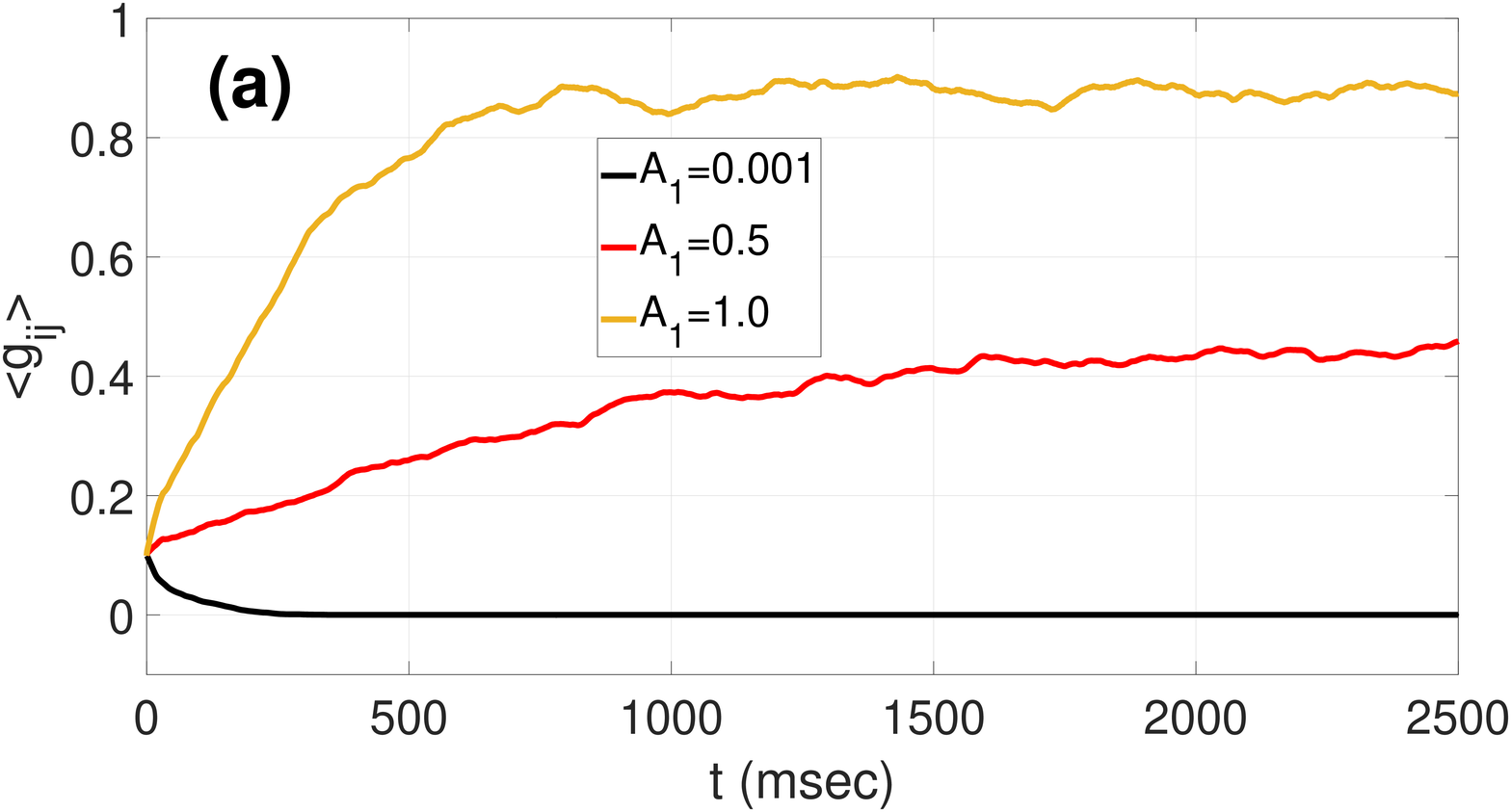}
\includegraphics[width=8.0cm,height=4.0cm]{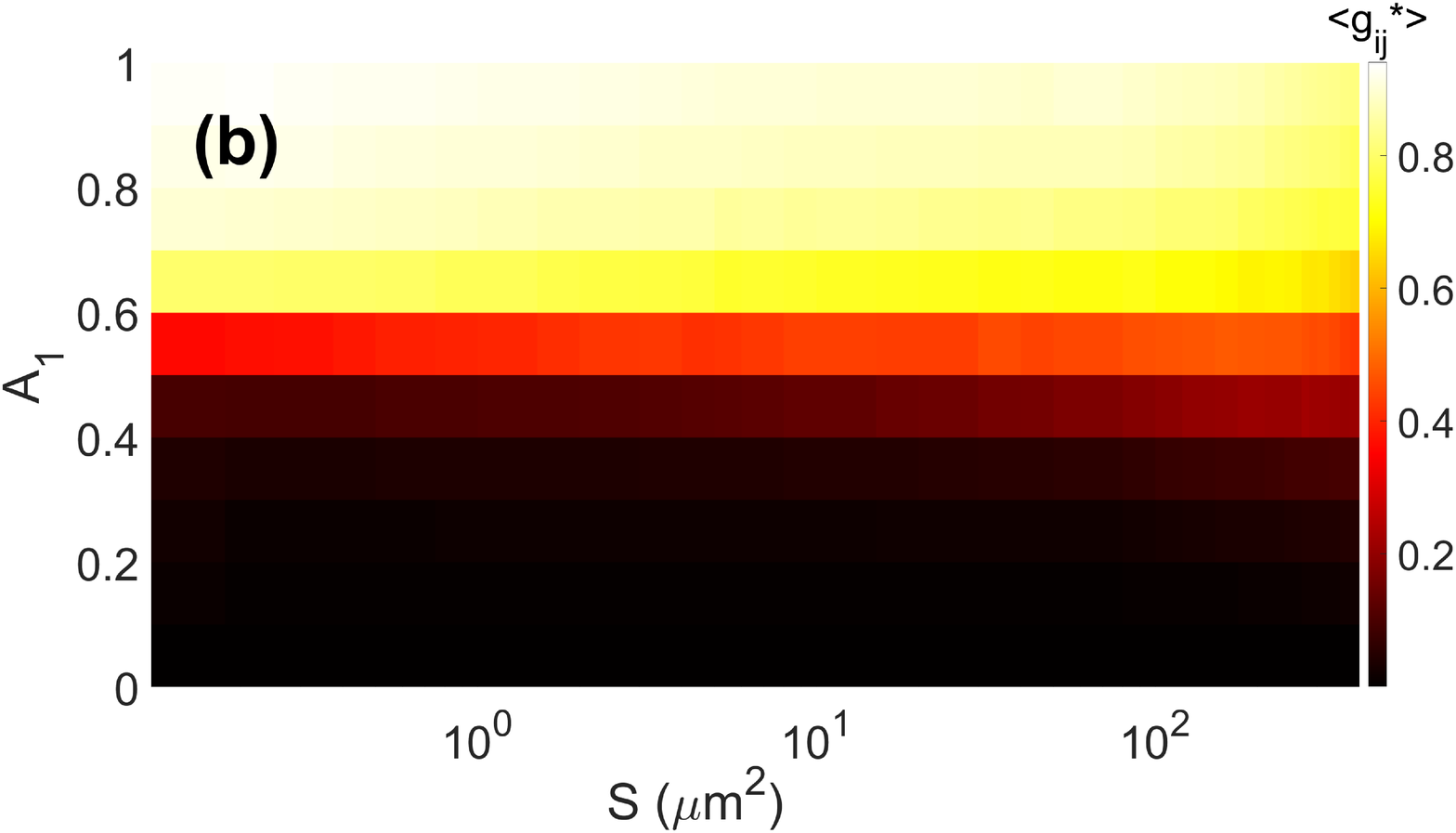}
\caption{
Effects of STDP on the population-averaged synaptic
weights $\langle g_{ij}\rangle$. Panel \textbf{(a)}: Time-evolution of $\langle g_{ij}\rangle$ at $S=200$.  Panel \textbf{(b)}: Variation of the  saturated population-averaged synaptic weight $\langle g_{ij}^*\rangle=\langle g_{ij}\rangle(t=2500)$ w.r.t. $A_1$ and $S$. 
$\langle k \rangle=9$,  $\beta=0.9$, $\tau_1= 20.0$, $A_2=0.5$, $\tau_2= 20.0$. }

\label{fig:7}
\end{figure}

%%%%%%%%%%%%%%%%%%%%%%%%%%%%%%%%%%%%%%%%%%%%%%%%%%%%%%%%%%%%%%%
Fig.~\ref{fig:7} shows the effects of STDP on the population-averaged synaptic weights $\langle g_{ij}\rangle$. In Fig.~\ref{fig:7}\textbf{(a)}, we show the time-evolution of $\langle g_{ij}\rangle$ for different values of $A_1$. All potentiations and depressions saturate at $\langle g_{ij}^*\rangle$ for $t\geq2000$ $\mathrm{msec}$ and the strongest potentiation and depression occur at $A_1=1$ and $A_1=0.001$, respectively. Fig.~\ref{fig:7}\textbf{(b)} shows the variation of $A_1$ with $S$ and the saturated population-averaged synaptic weight $\langle g_{ij}^*\rangle=\langle g_{ij}\rangle(t=2500)$. At each value of $A_1$, changing $S$ does not significantly change the potentiation or depression of $\langle g_{ij}^*\rangle$.

\begin{figure*}
\centering
\includegraphics[width=7.0cm,height=4.0cm]{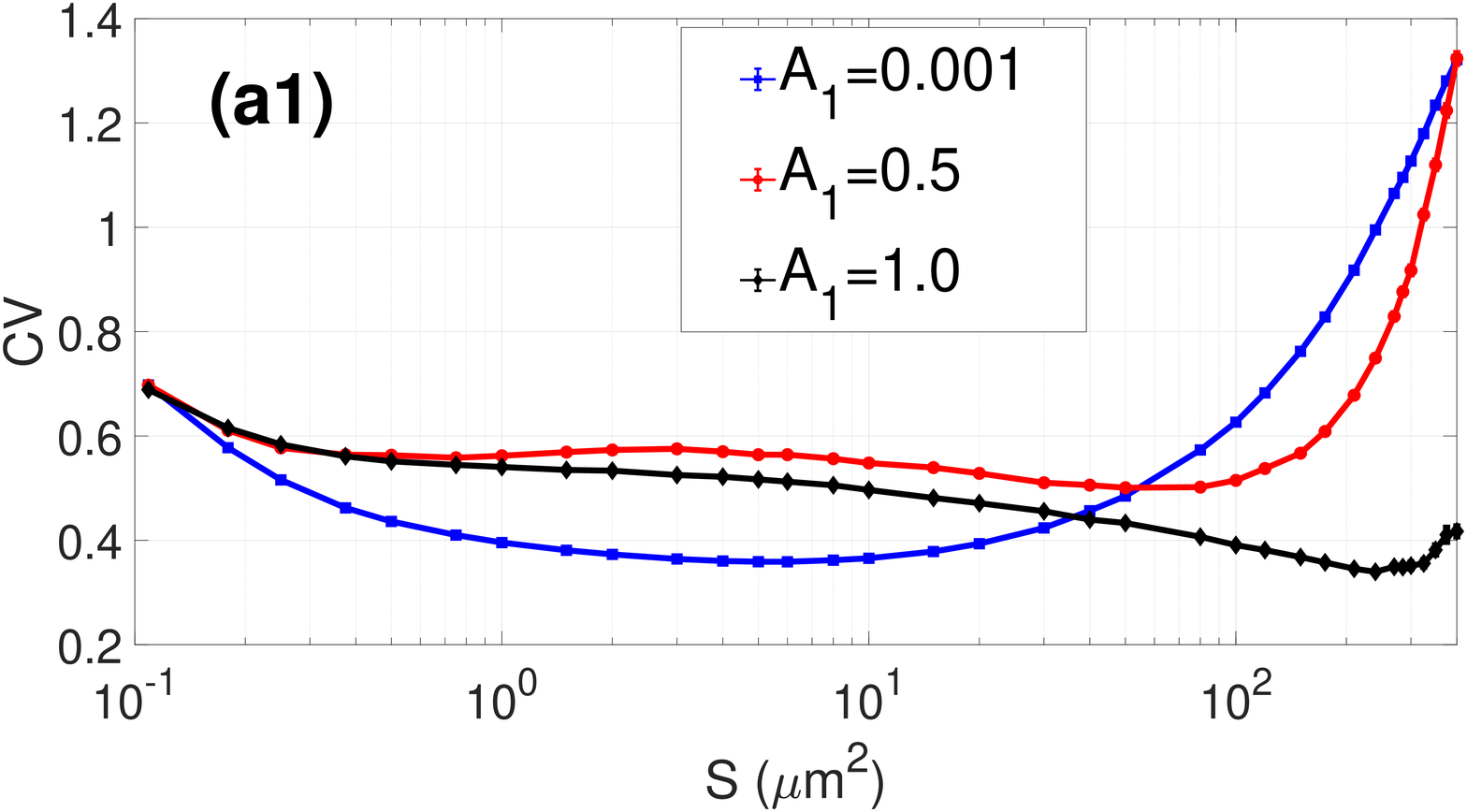}\includegraphics[width=7.0cm,height=4.0cm]{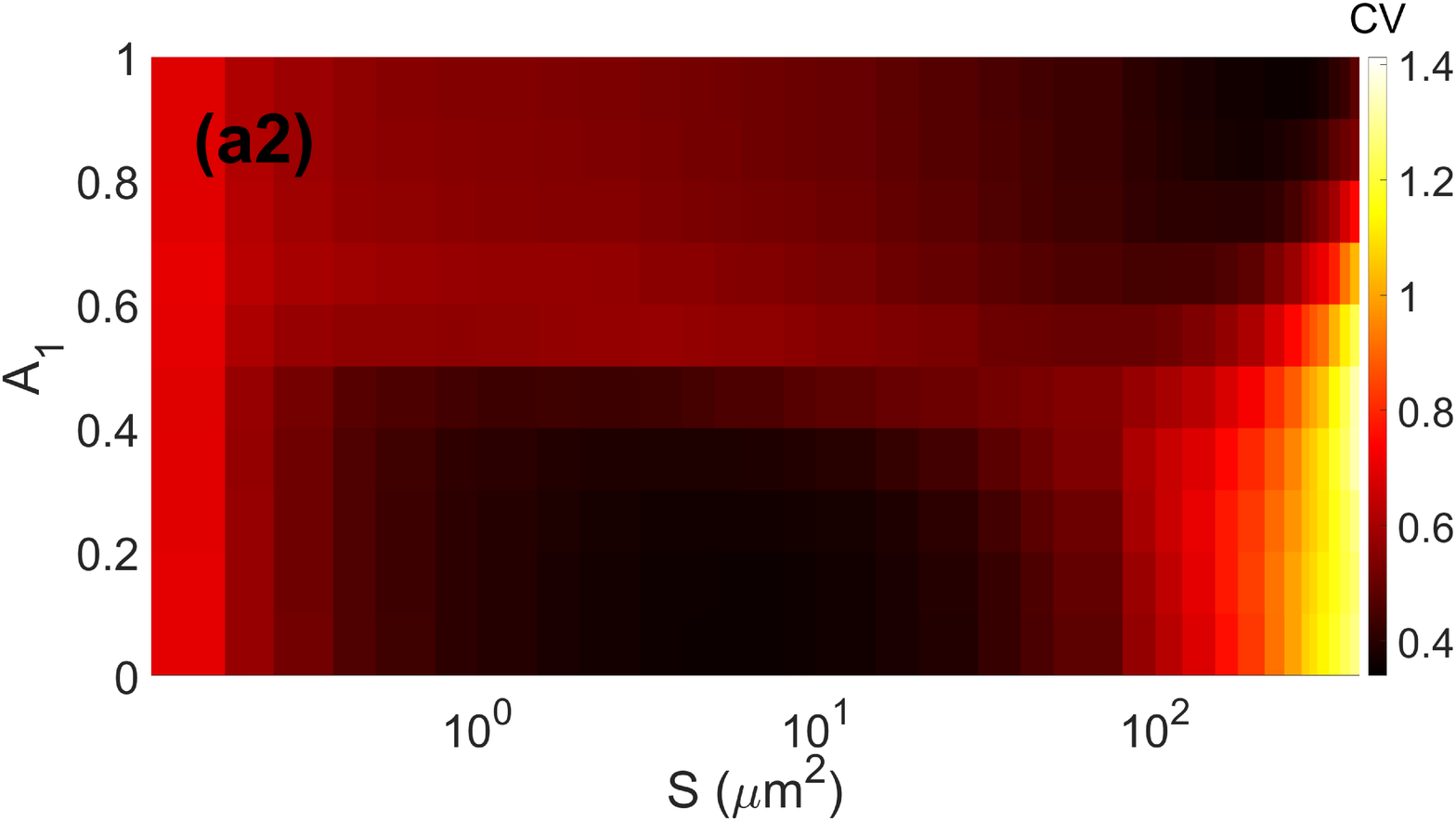}
\includegraphics[width=7.0cm,height=4.0cm]{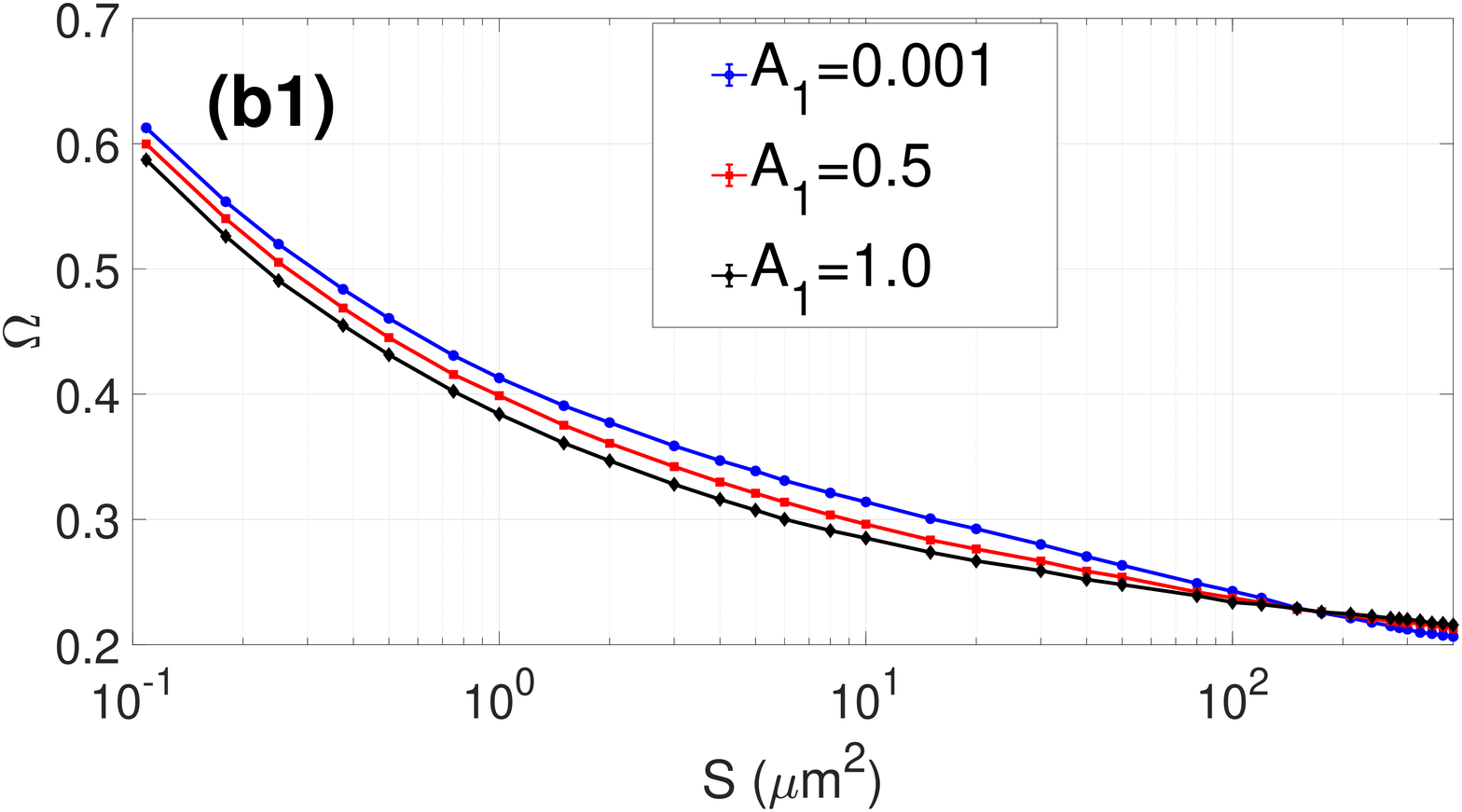}\includegraphics[width=7.0cm,height=4.0cm]{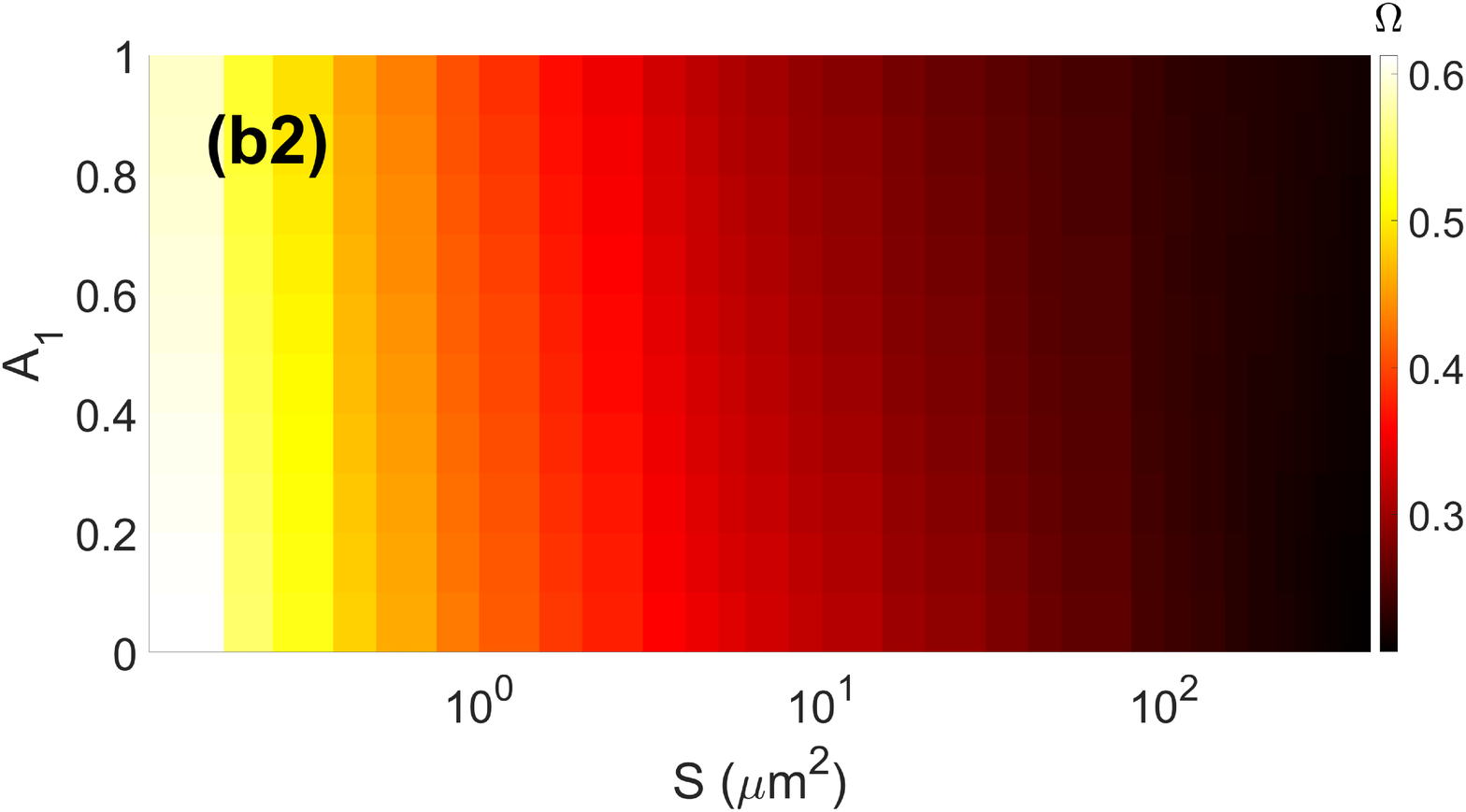}
\caption{
Variation of $\mathrm{CV}$ and $\Omega$ w.r.t. $S$ and the potentiation adjusting rate parameter $A_1$. In panels \textbf{(a1)} and \textbf{(a2)}, lower and higher values of $A_1$ enhance CR, while intermediate values destroy CR. In panels \textbf{(b1)} and \textbf{(b2)}, larger $S$ increases the degree of SS, especially at smaller $A_1$.  $\beta=0.9$, $\langle k \rangle=9$, $A_2=0.5$,  $\tau_1= 20.0$,  $\tau_2 = 20.0$.}

\label{fig:8}
\end{figure*}

 In Figs.~\ref{fig:8}\textbf{(a1)} and \textbf{(a2)}, the $\mathrm{CV}$ curves show a non-monotonic behavior as $S$ increases. In Fig.~\ref{fig:8}\textbf{(a1)}, we observe that (i) as $A_1$ increases, the resonant intermediate value of $S$ is shifted to the right, (ii) the best degrees of CR at small and large values of $A_1$ (which, according to Fig.~\ref{fig:7}, correspond to synaptic weakening and strengthening, respectively) are approximately the same, and (iii) at the mid-value of $A_1$ (i.e, $A_1=0.5$) where the synaptic coupling has potentiated to an intermediate strength (see, e.g., the red curve Fig.~\ref{fig:7}\textbf{(a)}), the minimum value of $\mathrm{CV}$ is higher.
 
Figs.~\ref{fig:8}\textbf{(b1)} and \textbf{(b2)} show a monotonic behavior in the degree of SS  as $S$ increases. We see, especially in  Fig.~\ref{fig:8}\textbf{(b1))}, that at larger values of $A_1$ and smaller values of $S$ (see, e.g., the black curve in Fig.~\ref{fig:8}\textbf{(b1))}, the degree of SS is better than at the smaller values of $A_1$ (see e.g., the red and blue curves in Fig.~\ref{fig:8}\textbf{(b1)}). However, at larger values of $S$  (i.e., $S\geq100$), the scenario reverses: the black curve goes above the red and the blue curves, meaning that at larger $S$ and smaller $A_1$ (i.e., weaker synapses) the network synchronizes better.

\begin{figure}
\centering
\includegraphics[width=8.0cm,height=4.0cm]{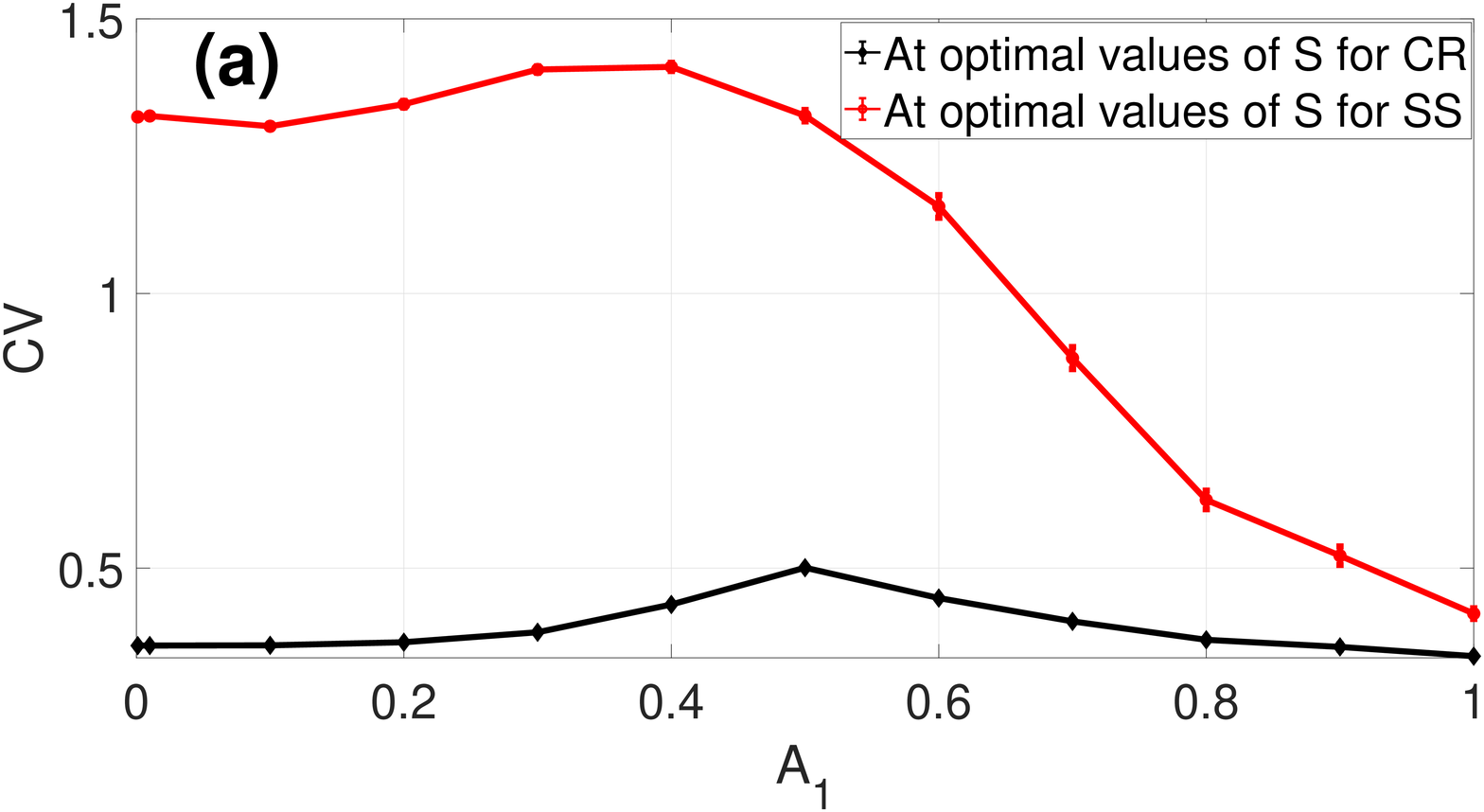}
\includegraphics[width=8.0cm,height=4.0cm]{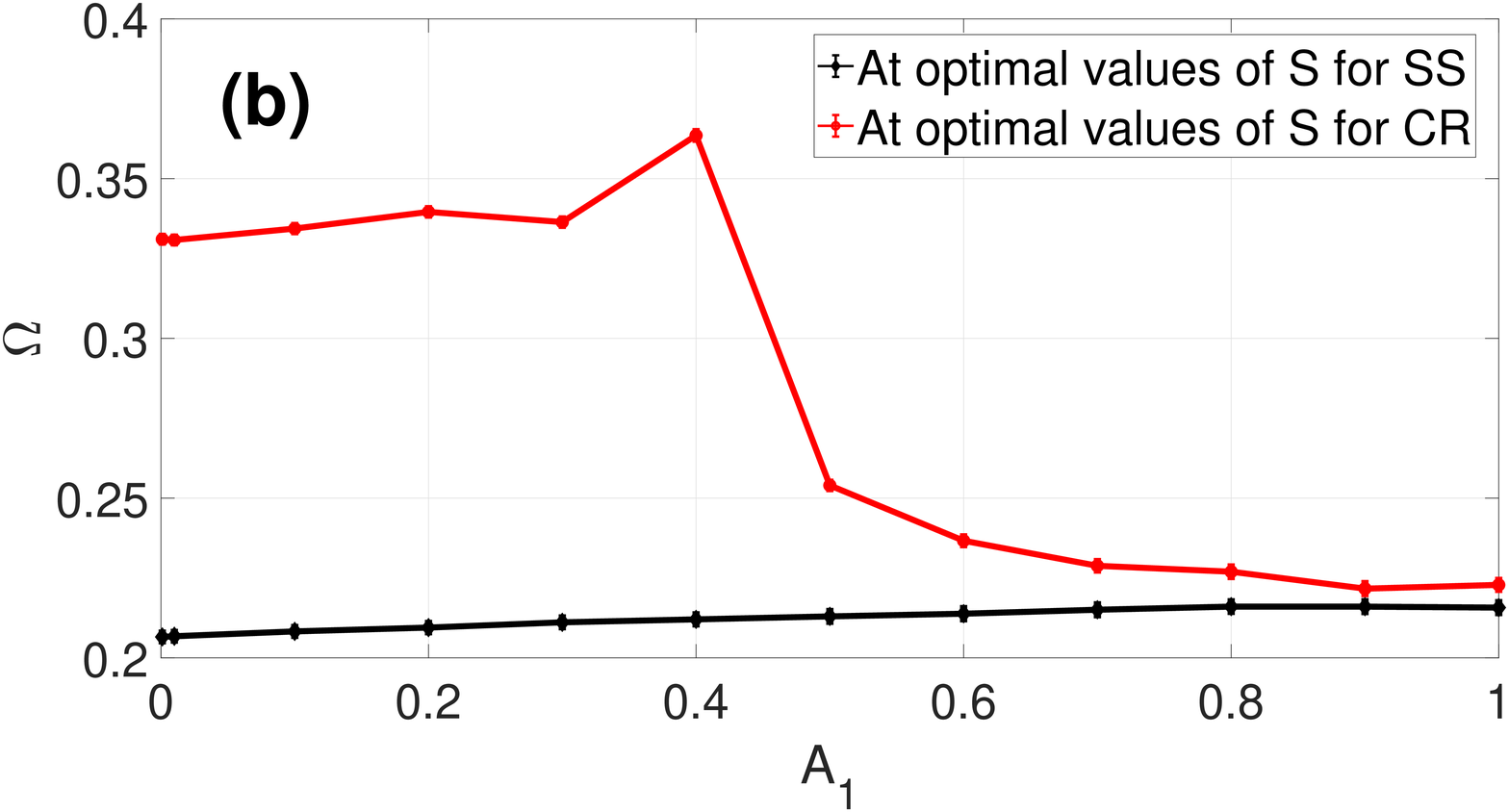}
\caption{
Variation of $\mathrm{CV}$ (panel \textbf{(a)}) and $\Omega$ (panel \textbf{(b)}) w.r.t. the potentiation adjusting rate $A_1$ at their optimal and non-optimal membrane patch area $S$. $\beta=0.9$, $\langle k \rangle=9$, $A_2=0.5$,  $\tau_1= 20.0$,  $\tau_2 = 20.0$.}.

\label{fig:9}
\end{figure}

In Fig.~\ref{fig:9}\textbf{(a)}, the black curve shows a monotonic increase (with a peak at $A_1=0.5$) and then decrease in the degree of CR as $A_1$ increases at resonant values of $S$.  The red curve shows that at the largest value of $S$, the degree of the CR can only be enhanced by increasing the value of $A_1$ (i.e., by strengthening the synapses). In Fig.~\ref{fig:9}\textbf{(b)}, the black curve shows a slight decrease in the degree of SS as $A_1$ increases. The red curve shows that at the resonant values of $S$, one can only enhance the degree of SS by increasing $A_1$. Therefore, by fixing the parameters at $\beta=0.9$, $\langle k \rangle=9$, $A_2=0.5$,  $\tau_1= 20.0$,  $\tau_2 = 20.0$, we can simultaneously enhance the degrees of CR and SS by choosing larger values of $A_1$. Thus, in the sequel we fix the adjusting rate parameter at $A_1=1$.

%%%%%%%%%%%%%%%%%%%%%%%%%%%%%%%%%%%%%%%%%%%%%%%%%%%%%%%%%%%%%%%
\subsubsection{ Depression temporal window $\tau_2$}
\begin{figure}
\centering
\includegraphics[width=8.0cm,height=4.0cm]{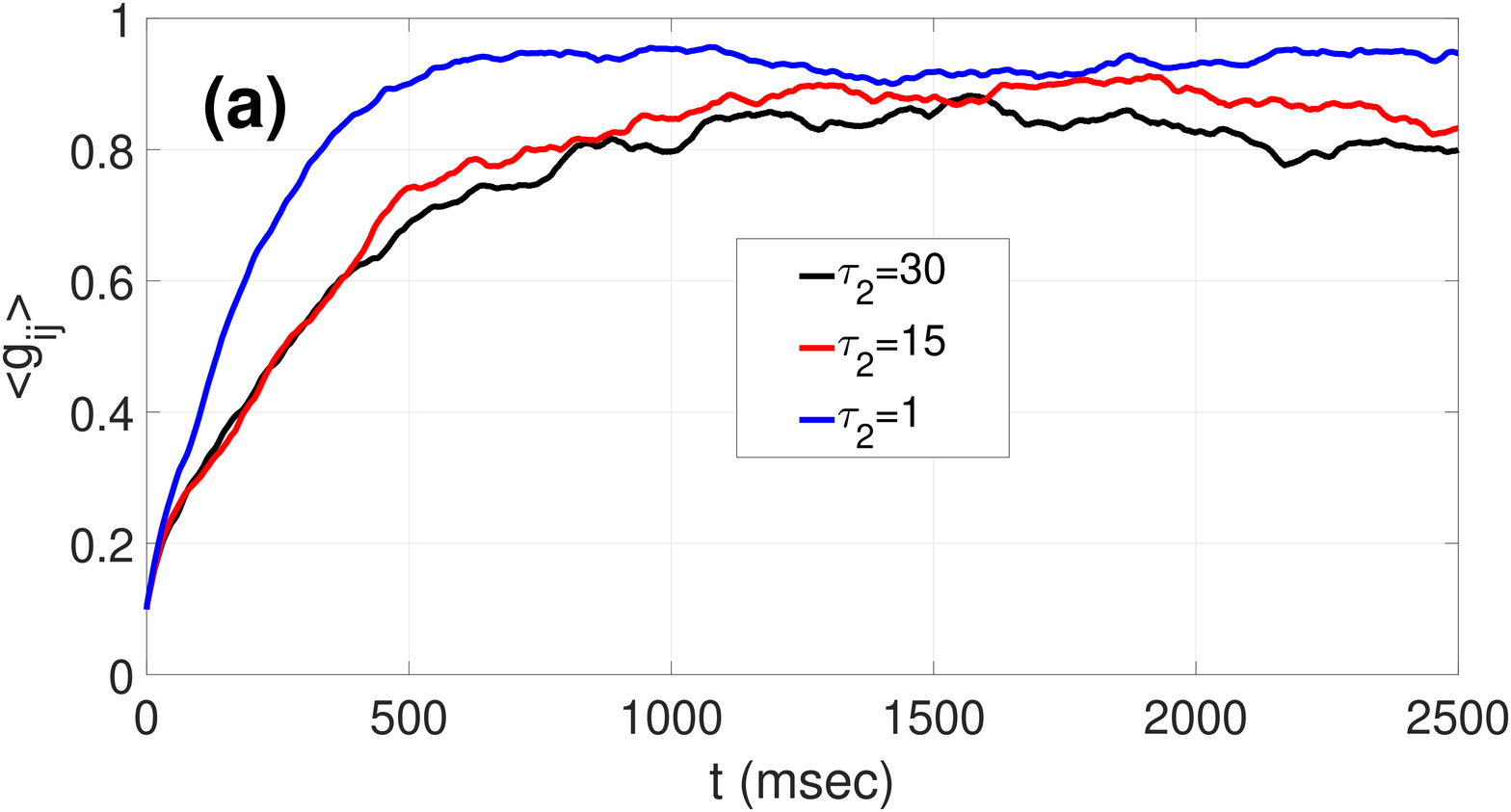}
\includegraphics[width=8.0cm,height=4.0cm]{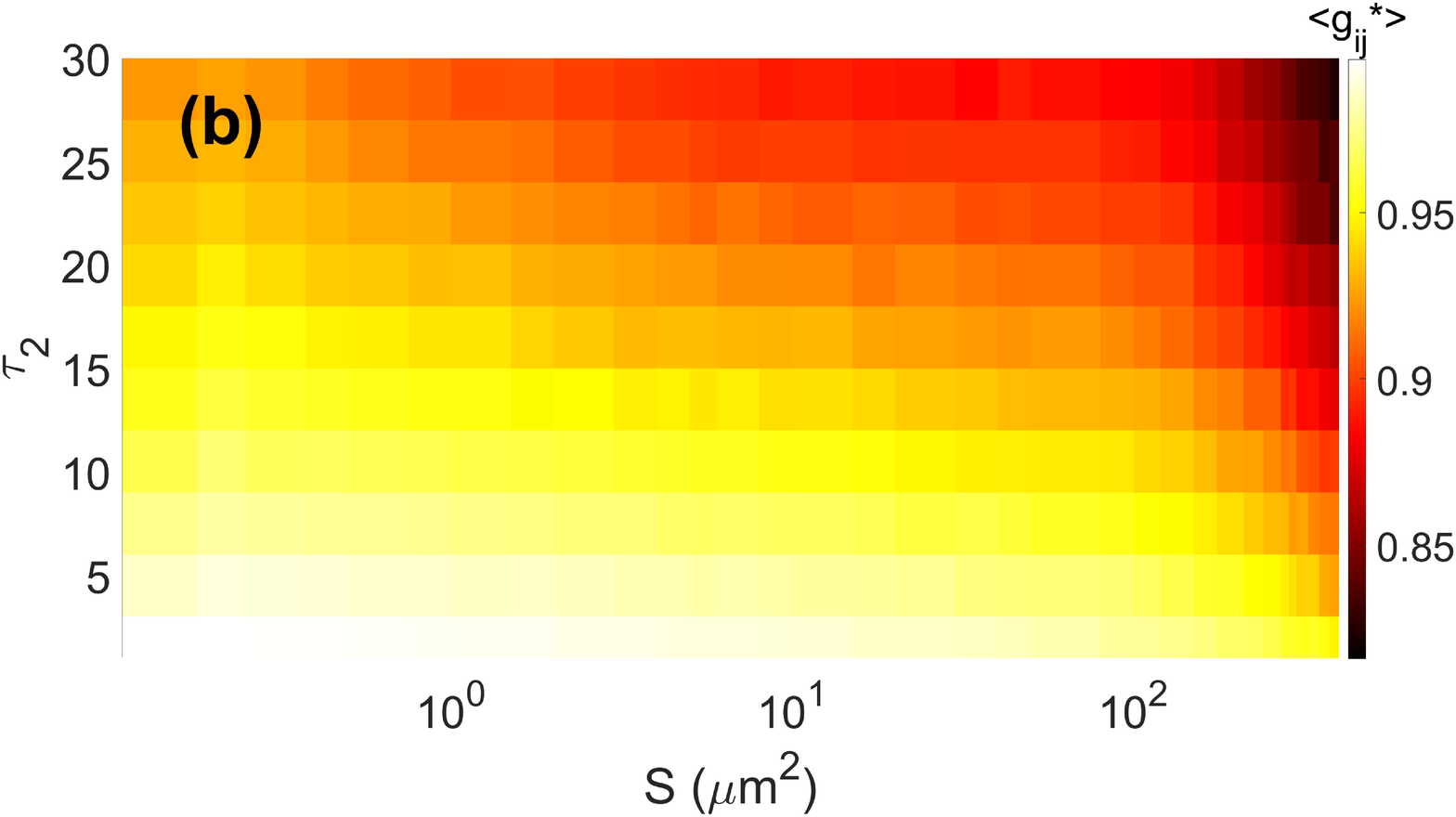}
\caption{
Effects of STDP on the population-averaged synaptic
weights $\langle g_{ij}\rangle$. Panel \textbf{(a)}: Time-evolution of $\langle g_{ij}\rangle$ at $S=400$.  Panel \textbf{(b)}: Variation of the  saturated population-averaged synaptic weight $\langle g_{ij}^*\rangle=\langle g_{ij}\rangle(t=2500)$ w.r.t. the depression temporal window $\tau_2$ and $S$. 
$\beta=0.9$, $\langle k \rangle=9$, $A_1=1.0$, $A_2=0.5$,  $\tau_1 = 20.0$. }
\label{fig:10}
\end{figure}

%%%%%%%%%%%%%%%%%%%%%%%%%%%%%%%%%%%%%%%%%%%%%%%%%%%%%%%%%%%%%%%
In Fig.~\ref{fig:10}\textbf{(a)}, we show the time-evolution of $\langle g_{ij}\rangle$ for different values of the temporal window parameter $\tau_2$ at $S=400$. We observe that all values of $\tau_2$ potentiate the synapses with the strongest potentiation occurring at the lower values of $\tau_2$ (since with $A_1=1.0$,  $\tau_1=20.0$, and $A_2=0.5$, we have $A_1\tau_1>A_2\tau_2$, for all $\tau_2\in(0,30]$).  Fig.~\ref{fig:10}\textbf{(b)} shows the variation of $\tau_2$ with $S$ and the saturated population-averaged synaptic weight $\langle g_{ij}^*\rangle$. At each value of $\tau_2$, as $S$ increases, the weaker $\langle g_{ij}^*\rangle$ becomes. However, $\langle g_{ij}^*\rangle$ never weakens below the mean value of the initial distribution.

\begin{figure*}
\centering
\includegraphics[width=7.0cm,height=4.0cm]{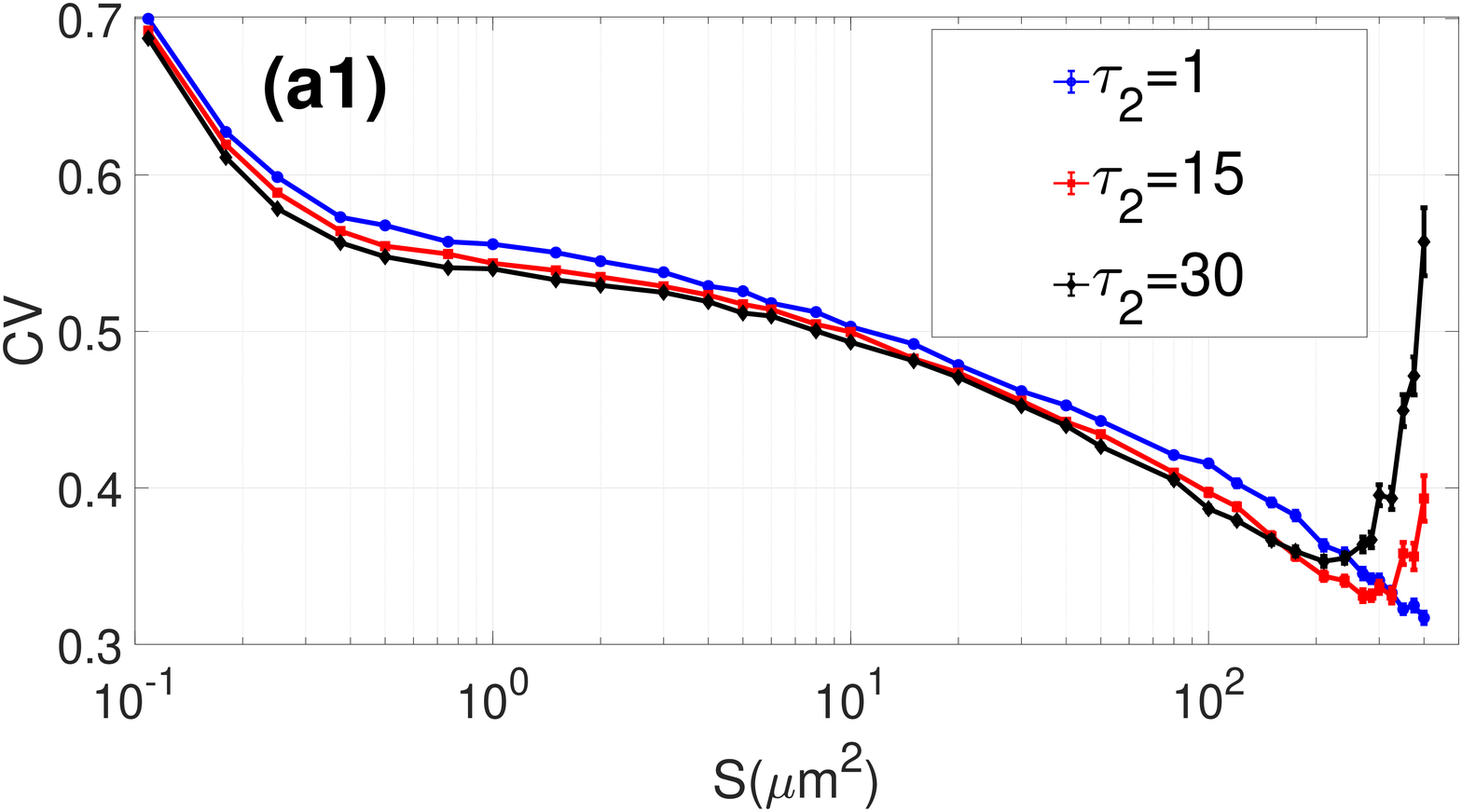}\includegraphics[width=7.0cm,height=4.0cm]{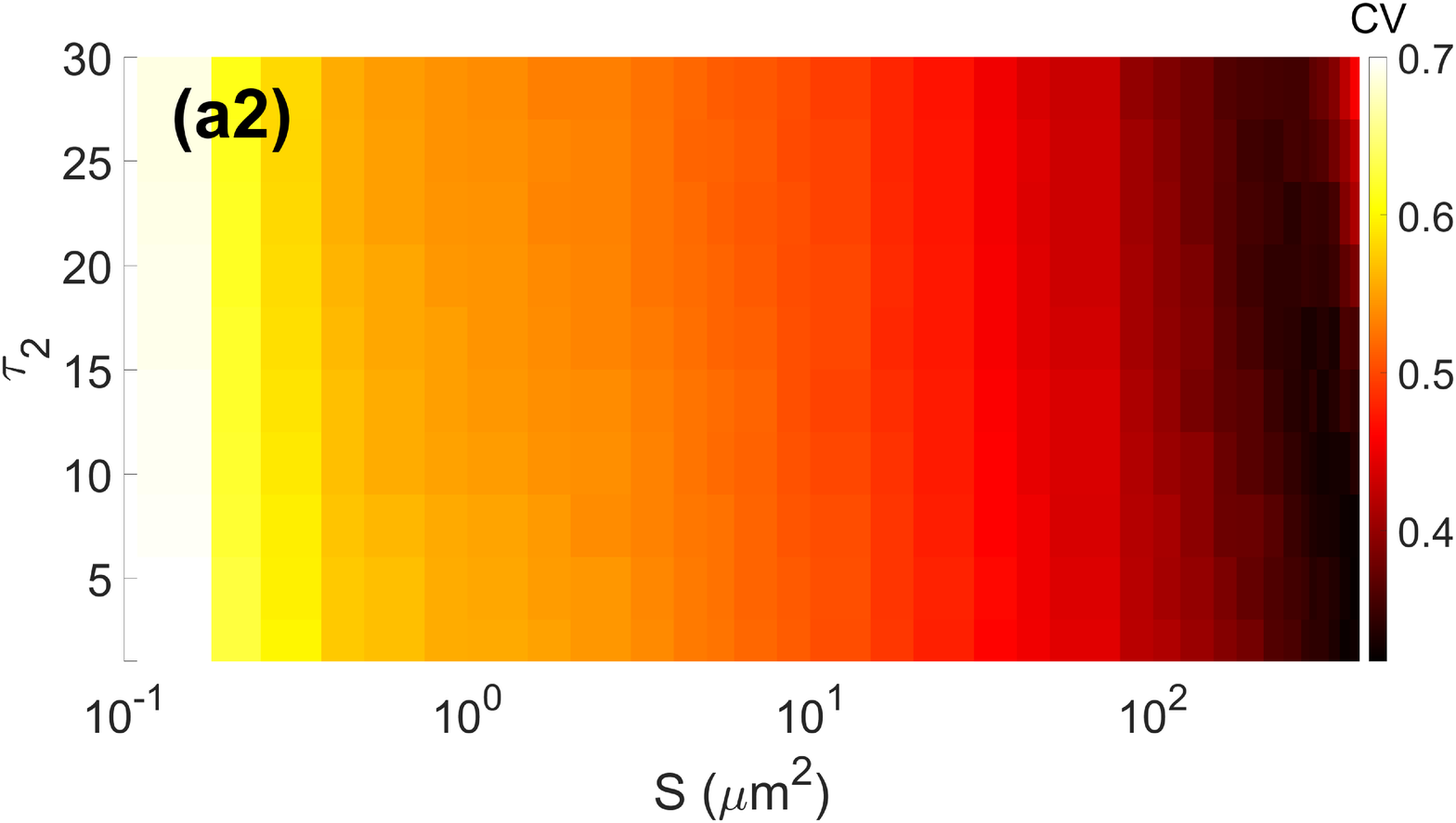}
\includegraphics[width=7.0cm,height=4.0cm]{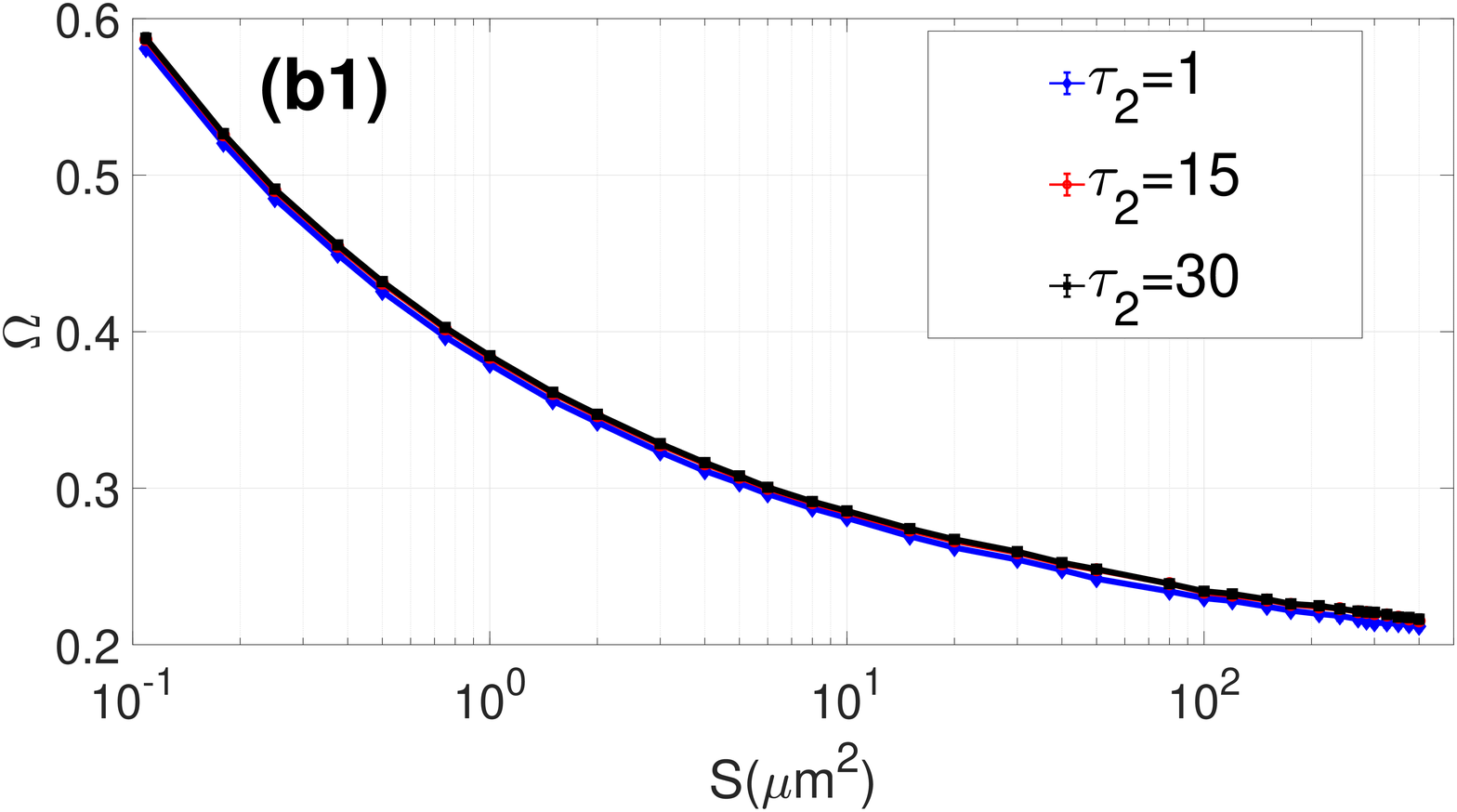}\includegraphics[width=7.0cm,height=4.0cm]{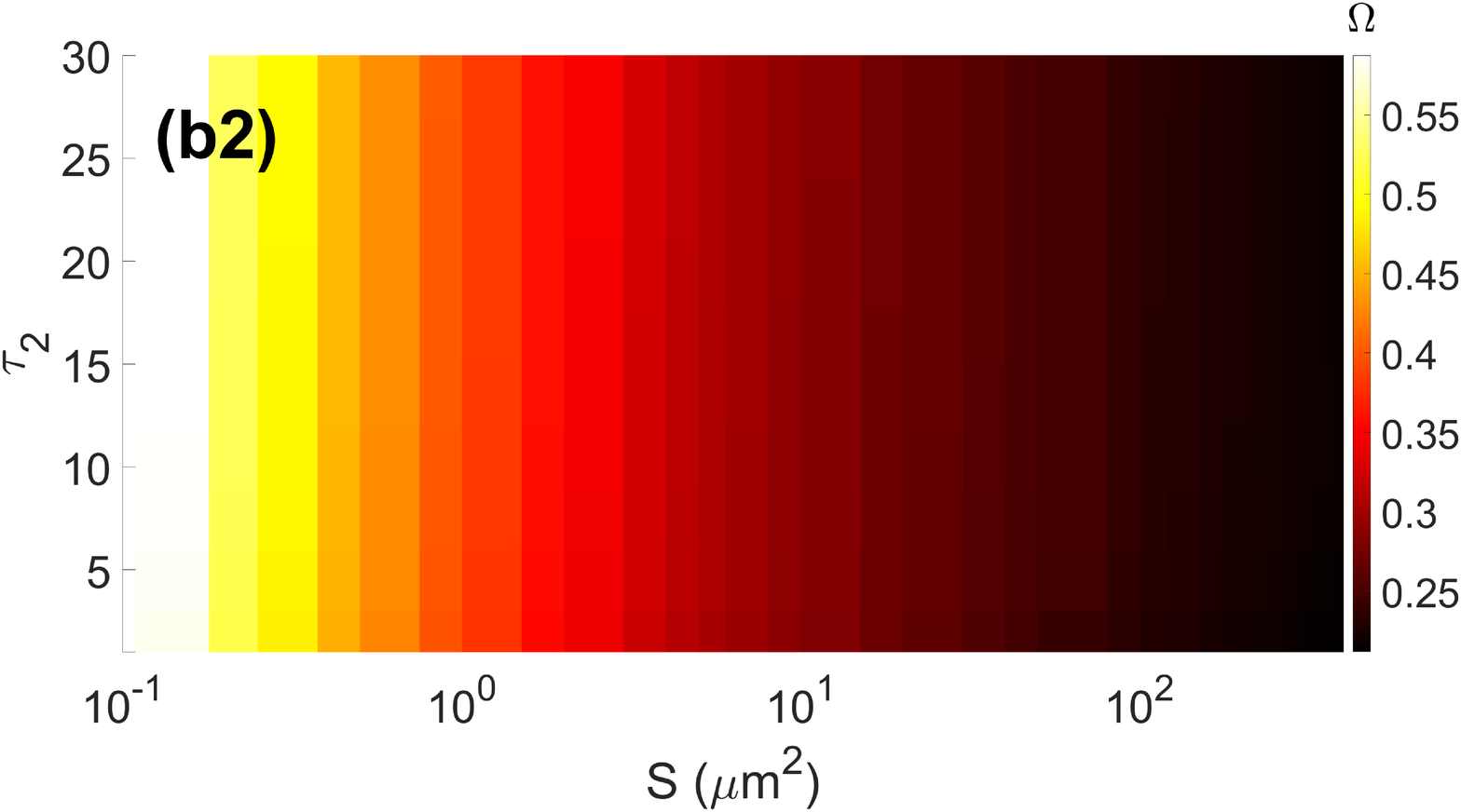}
\caption{Variation of $\mathrm{CV}$ and $\Omega$ w.r.t. $S$ and the depression temporal window $\tau_2$. Panels \textbf{(a1)} and \textbf{(a2)} show a non-monotonic and monotonic behavior (depending on the value of $\tau_2$) of the $\mathrm{CV}$ curves as $S$ increases. 
Panels \textbf{(b1)} and \textbf{(b2)} show a monotonic decrease of the $\Omega$ curves as $S$ increases. $\beta=0.9$, $\langle k \rangle=9$, $A_1=1.0$, $A_2=0.5$,  $\tau_1= 20.0$.}
\label{fig:11}
\end{figure*}

Figs.~\ref{fig:11}\textbf{(a1)} and \textbf{(a2)} show the variation of the $\mathrm{CV}$ (for different values of $\tau_2$) as $S$ increases. When $\tau_2\geq12$ the $\mathrm{CV}$ curves show a non-monotonic behavior (see, e.g., the black and red curves in Fig.~\ref{fig:11}\textbf{(a1)}). But, when $\tau_2<12$ (see, e.g., the blue curve in Fig.~\ref{fig:11}\textbf{(a1)}), the $\mathrm{CV}$ curves decreases monotonically as $S$ increases within the interval of simulation (i.e., $S\in[0.10,400]$). Moreover, we observe that when $S\in[0.10,100]$, larger $\tau_2$ induce a better degree of CR than smaller values. But for $S\in(100,400]$ the scenario reverses, indicating a poorer degree of CR. Figs.~\ref{fig:11}\textbf{(b1)} and \textbf{(b2)} show the monotonic increase in the degree of SS (for different values of $\tau_2$) as $S$ increases. We observe that increasing $\tau_2$ essentially leaves the degree of SS unchanged. 

\begin{figure}
\centering
\includegraphics[width=8.0cm,height=4.0cm]{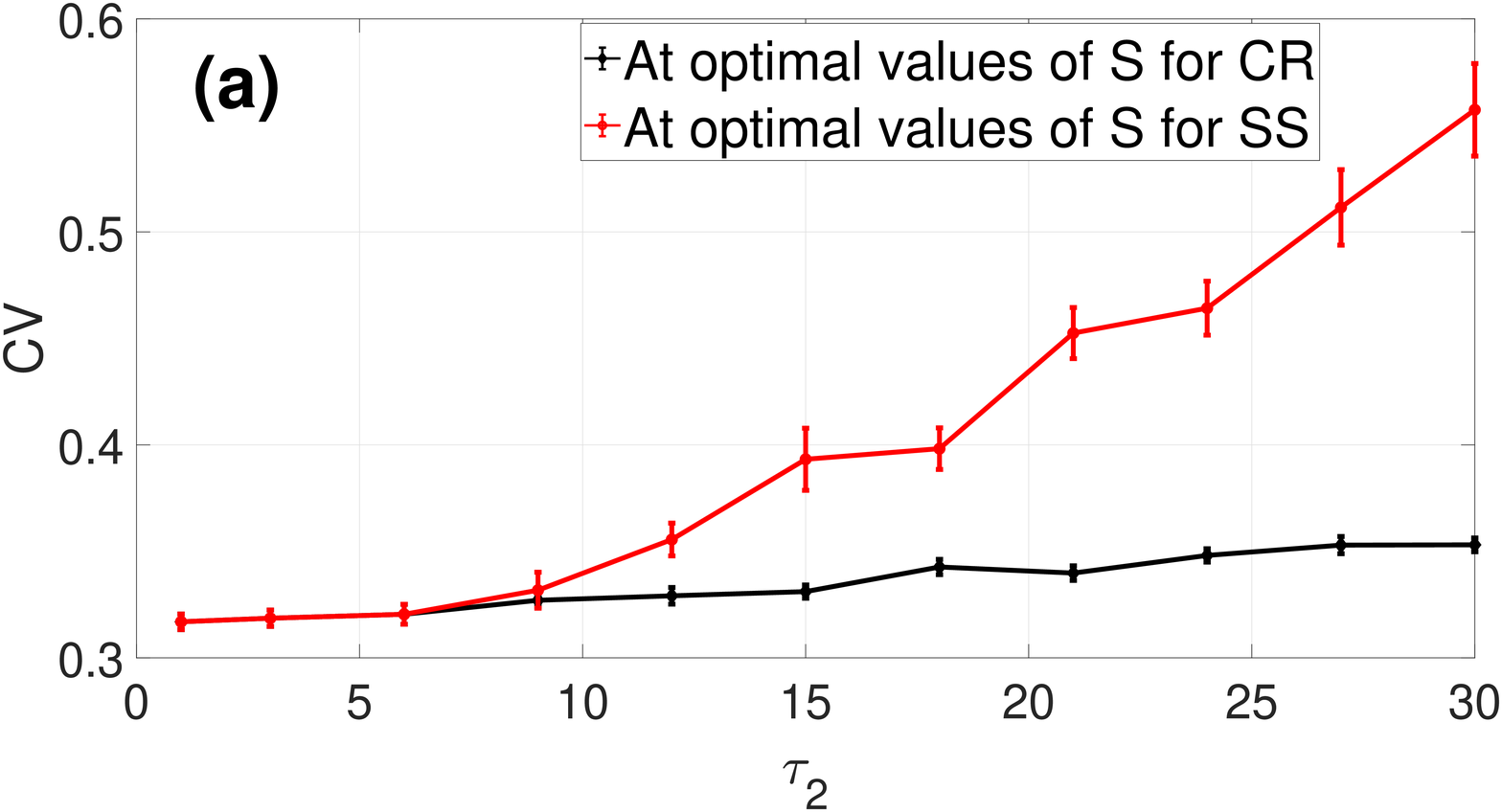}
\includegraphics[width=8.0cm,height=4.0cm]{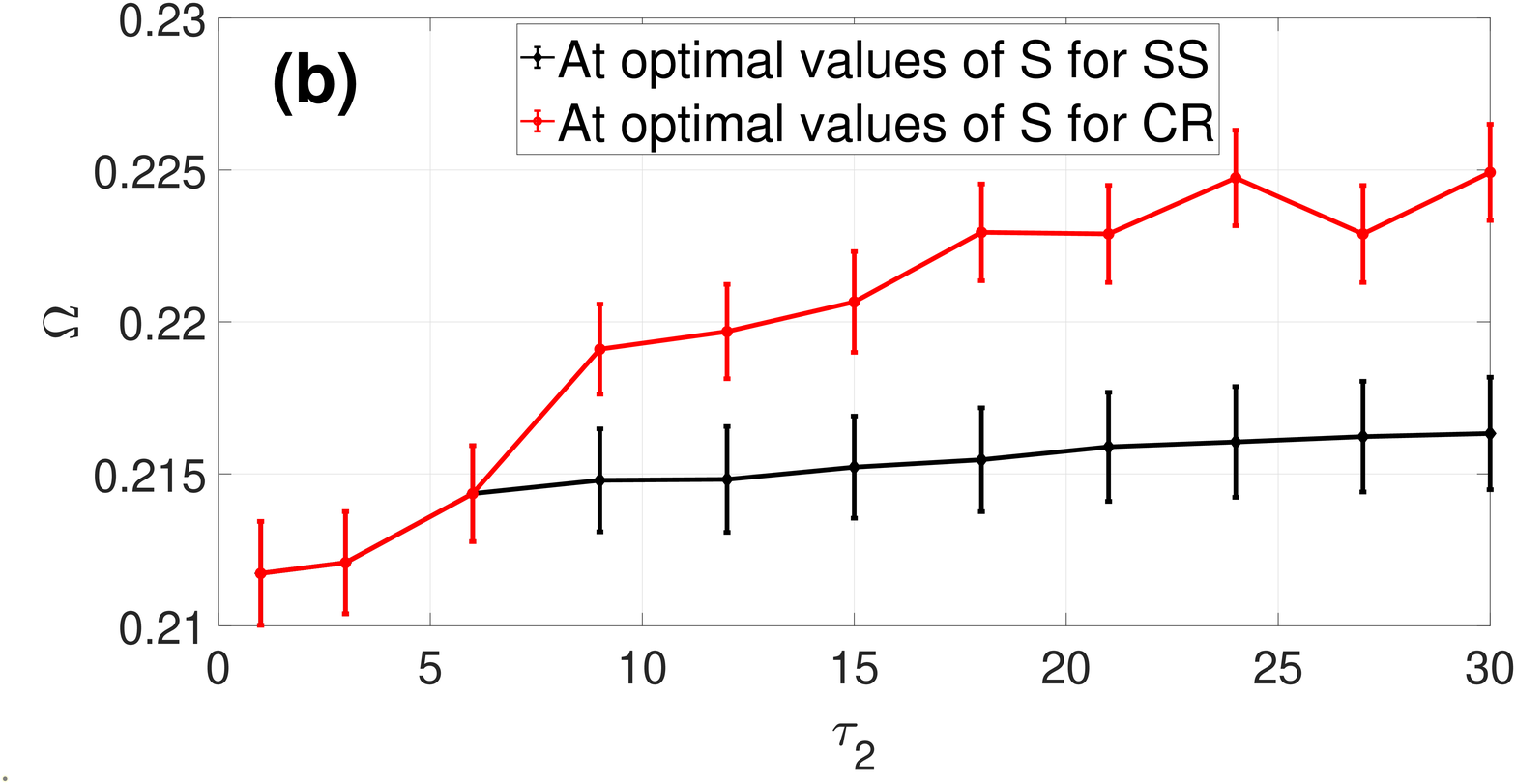}
\caption{Variation of $\mathrm{CV}$ (panel \textbf{(a)}) and $\Omega$ (panel \textbf{(b)}) w.r.t. the depression temporal window $\tau_2$  at their optimal and non-optimal membrane patch area $S$. $\beta=0.9$, $\langle k \rangle=9$, $A_1=1.0$, $A_2=0.5$, $\tau_1= 20.0$.}
\label{fig:12}
\end{figure}

In Fig.~\ref{fig:12}\textbf{(a)}, the black curve shows a monotonic but gentle decrease in the degree of CR as $\tau_2$ increases at the corresponding resonant values of $S$. In Fig.~\ref{fig:12}\textbf{(b)}, the black curve also shows that the degree of SS also decreases with $\tau_2$ at the largest value of the $S$. The red curves in Figs.~\ref{fig:12}\textbf{(a)} and \textbf{(b)} show that at optimal values of $S$ increasing $\tau_2$ deteriorates both CR and SS.

One would intuitively expect that  the weak noise limit (i.e., as $S$ becomes larger) leads to a Poisson
statistics of the mean inter-spike intervals. But this is not always the case in the results presented. To explain this observation, first, we note that membrane patch area $S$ is a finite physical quantity with the largest value of $S=400$ in our calculations. It is important to note that the value of $\mathrm{CV}$ depends not only on the value of $S$ (which controls the noise intensity) but also significantly depends on the values of the other parameters. For example, in Fig.~\ref{fig:3}\textbf{(a1)}, at the largest value of $S$ and for the average degree $\langle k \rangle =2 $, the value of CV may be even more variable than the Poisson process (i.e., $\mathrm{CV}>1$), close to the Poisson process (i.e.,  $\mathrm{CV}\approx1$) for $\langle k \rangle =25$, or even less variable than a Poisson process (i.e, $\mathrm{CV}<1$) for $\langle k \rangle =9$. See also  Fig.~\ref{fig:11}, with the depression temporal window $\tau_{_{2}}$. Thus, $S$ is not the only element that influences the regularity of the spiking.

%%%%%%%%%%%%%%%%%%%%%%%%%%%%%%%%%%%%%%%%%%%%%%%%%%%%%%%%%%%%%%%
\section{Conclusions and perspectives}\label{Sec. V}
%%%%%%%%%%%%%%%%%%%%%%%%%%%%%%%%%%%%%%%%%%%%%%%%%%%%%%%%%%%%%%%
In summary, we have numerically investigated the phenomena of coherence resonance (CR), stochastic synchronization (SR), and their interplay in a Watts-Strogatz small-world network of excitable Hodgkin-Huxley neurons driven by channel noise in the presence of spike-timing-dependent plasticity (STDP). The numerical results indicate that (i) intermediate values of membrane patch area $S$ (which controls the intensity of the channel noise) induce the best degree of temporal coherence of the spiking activity via CR, and (ii) large values of the membrane patch area (which correspond to small values of the noise intensity of the channel noise) induce the best degree of SS.

It is demonstrated that the network parameters, i.e., the average degree $\langle k \rangle$ and the rewiring probability $\beta$, play a significant role in the degrees of CR and the interplay between CR and SS. For the average degree $\langle k \rangle$, it is shown that (i) at the resonant (intermediate) value of $S$ (where the degree of CR is the best), increasing $\langle k \rangle$ in the interval $[1,25]$ decreases the degree of CR, (ii) at the optimal (largest) value of $S$ (where the degree of SS is the best), increasing $\langle k \rangle$ increases the degree of SS, (iii) at the optimal value of $S$ for SS, the poor degree of CR can be enhanced by increasing $\langle k \rangle$ in the interval $[1,25]$ up to the resonant value of $\langle k \rangle=9$ where the best degree CR occurs, and (iv) at resonant values $S$ for CR, the relatively poor degree of SS can be enhanced by increasing $\langle k \rangle$.  For the rewiring probability $\beta$, it is shown that (i) at the resonant values of $S$ for CR, increasing the randomness of the network connectivity (i.e., increasing $\beta$) slowly decreases the degree of CR, (ii) at the optimal value $S$ for SS, increasing $\beta$ increases the degree of CR, (iii) at both resonant values of $S$ and the largest value of $S$, the high degree of SS remains essentially the same.

It is also shown that the potentiation adjusting rate $A_1$ plays a more significant role in the degrees of CR, SS, and on their interplay, compared to the depression temporal window $\tau_2$.
For the adjusting rate parameter $A_1$, it is shown that (i) at the resonant value of $S$ for CR, increasing $A_1$ in (a) the interval $[0.001,0.5)$, decreases the degree of CR and (b) the interval $(0.5,1.0]$ increases the degree of CR, (ii) at the optimal value of $S$ for SS, increasing $A_1\in[0.001,1.0]$ slowly decreases the degree of SS, (iii) at the optimal value of $S$ for SS, the relatively poor degree of CR can be enhanced by increasing $A_1\in[0.001,1.0]$, 
and (iv) at resonant values $S$ for CR, the relatively poor degree of SS can also be enhanced by increasing $A_1$ in the same interval. For the temporal window parameter $\tau_2$, it is shown that (i) at the resonant values of $S$ for CR, increasing $\tau_2$ in the interval $[1,30]$ slowly decreases the degree of CR, whereas it leave almost the degree of SS unchanged, and (ii) at the optimal value $S$ for SS, increasing $\tau_2\in[1,30]$ rapidly decreases the degree of CR, whereas it slowly decreases the degree of SS. Hence, decreasing the value of $\tau_2$ would enhance the degree of CR and leaves the high degree of SS essentially constant.

In conclusion, the phenomena of coherence resonance and stochastic synchronization depend on the parameters governing the network topology, the STDP learning rule, and the channel noise which can combine in various ways to concurrently (or separately) enhance the degrees of CR and SS. Our results could shed some light on the efficient coding mechanisms based on the spatio-temporal coherence of spike trains in neural networks and their information processing capability. The interplay between CR and SS when the network topology is time-varying is interesting and deserves further investigations.

It is well-known that information processing in the human neural system is metabolically expensive. Although the brain is only 2\% of the body's weight, it accounts for up 20\% of its resting energy consumption \cite{attwell2001energy,laughlin2001energy}. Thus, the requirement for this significant metabolic energy has important implications for information processing in the brain. The availability of energy and its consumption could determine neural circuitry and activity patterns by favoring energetically efficient wiring patterns \cite{mitchison1992axonal}, and neural coding via phenomena such as coherence resonance and synchronization \cite{yamakou2016ratcheting,xie2022phase,huang2021energy,wu2020energy}. 
It has been shown that synapse connections can be created and activated in an adaptive way when field energy is exchanged between neurons, with significant consequences on their synchronization \cite{xie2022phase}. It has also been shown that the Hamilton energy of each neuron in a network can keep pace with other neurons when complete synchronization is stabilized within a finite transient period \cite{huang2021energy}.

But the flow of energy in a neural network exhibiting both coherence resonance and synchronization in the absence and presence of STDP is still lacking. Furthermore, it should be rewarding to understand this energy flow in neural networks under electromagnetic radiation, which is ubiquitous in neural systems and has been shown to affect synchronization patterns significantly \cite{ma2017synchronization}. Thus, an imperative future research direction on the interplay between coherence resonance and synchronization in neural networks driven by STDP will be to investigate the energy aspect of this interplay. This could be done by using the method from Helmholtz theorem \cite{kobe1986helmholtz,chun2016calculation} to calculate the Hamilton energy and its consumption in the STDP-driven neural network exhibiting both coherence resonance and synchronization. What parameters control the efficient flow and consumption of energy flow in adaptive memristive neural networks exhibiting both coherence resonance and synchronization? How do energy flow and electromagnetic radiation control synapses between neurons? What consequences does this have on the co-existence of coherence resonance and synchronization in a small-work and random neural network? These are bio-physically relevant questions to be addressed.

\section*{Acknowledgments}
MEY acknowledges support from the Deutsche Forschungsgemeinschaft (DFG, German Research Foundation) -- Project No. 456989199. Research at Perimeter Institute is supported in part by the Government of Canada through the Department of Innovation, Science and Economic Development Canada and by the Province of Ontario through the Ministry of Economic Development, Job Creation and Trade.

\section*{Data and Code availability statement} 
The simulation data that support the findings of this study are available within the article. The code used to obtain the results of this article is made publicly available here~\cite{stdp_code}.

\section*{Declaration of competing interest}
The authors declare that there is no conflict of interest in relation to this article.
%
%

% \bibliographystyle{unsrt}
% \bibliography{ref} %Produces the bibliography via BibTeX.
\providecommand{\noopsort}[1]{}\providecommand{\singleletter}[1]{#1}%

\end{document}